\documentclass[preprint,12pt]{elsarticle}
\usepackage{amssymb,amsmath}
\usepackage{graphicx}
\biboptions{sort&compress}

\journal{Communications in Nonlinear Science and Numerical Simulation}

\begin{document}
\begin{frontmatter}
\title{Delocalized periodic vibrations in nonlinear LC and LCR electrical chains}

\author{G.M.Chechin}\ead{gchechin@gmail.com}
\author{S.A.Shcherbinin}

\address{Department of Physics, Southern Federal University, Zorge 5, Rostov-on-Don 344090, Russia}

\begin{abstract}

We consider an electrical LC-chains of $N$ nonlinear capacitors coupled by linear inductors assuming that voltage dependence of capacitors represents an even function. We prove that only $5$ symmetry determined nonlinear normal modes (NNM) can exist in the considered system. The stability of all these dynamical regimes for different $N$ is studied with the aid of the group-theoretical method [Physical Review E 73 (2006) 36216] which allows to simplify radically the variational systems appearing in the Floquet stability analysis. The scailing of the voltage stability threshold in the thermodynamic limit $N\rightarrow\infty$ is determined for each NNM. We show that the above group-theoretical method can be used for studying NNMs and their stability in the LCR-chains containing not only inductors and capacitors, but also resistors and external sources of time-periodic voltage.
\end{abstract}
\begin{keyword}
nonlinear dynamics, lattice models, nonlinear normal modes, invariant manifolds, group-theoretical methods
\end{keyword}
\end{frontmatter}

\section{Introduction}
\label{Intro}
In recent years, studying of nonlinear vibrations in mesoscopic systems of different physical nature received much attention. In particular, discrete breathers and various soliton-like excitations were actively studied in cantilever arrays~\cite{kanti-1,kanti-2}, granular crystals~\cite{gran}, Josephson junction lattices~\cite{Jos-1,Jos-2}, photonic crystals~\cite{photonic}, electrical lines~\cite{iran1,iran2,iran3,Osting,English}, etc. It is essential that in contrast to crystals, where only indirect experiments are possible, in mesoscopic systems one can often observe different dynamical objects directly.

\begin{figure}[htb]
\begin{center}
\includegraphics{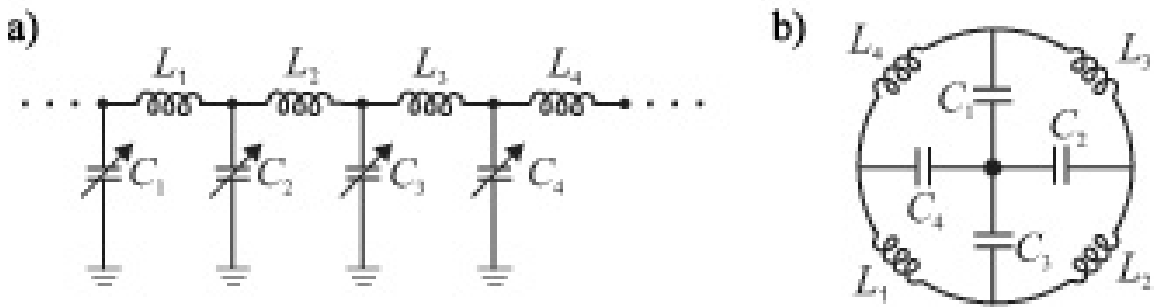}
\caption{Chain of nonlinear capacitors coupled by linear inductors (LC-chain)}\label{fig1}
\end{center}
\end{figure}

In Ref.~\cite{Osting}, the existence and linear stability of the so-called $\pi$-mode (zone boundary mode) in one-dimensional nonlinear electrical lattice were studied. This system, constructed on a silicon substrate by CMOS technology, represents a chain of nonlinear (voltage-dependent) capacitors coupled by linear inductors. This so called LC-chain is depicted schematically in Fig.\ref{fig1}a. Using periodic boundary conditions, one can imagine that the considered circuit represents a ring of $N$ cells. For the case $N=4$, such ring is depicted in Fig.\ref{fig1}b.

The dependence $C(V)$ of the capacity $C$ on the voltage $V$ can be rather complex, but the authors of the discussed reference used the following simple form of the function $C(V)$:
\begin{equation}\label{capacity}
C(V)=C_0(1-bV^2),
\end{equation}
where $C_0$ and $b$ are positive constants.

Applying the Kirchhoff laws for quasi-stationary current to the circuit in Fig.\ref{fig1}, one can obtain the following equations (see Ref.~\cite{Osting}):
\begin{equation}\label{kir}
   L_j\frac{dI_j}{dt}=V_j-V_{j+1},\; \frac{dQ(V_j)}{dt}=I_{j-1} - I_j.
\end{equation}
Here $Q(V_j)=Q(V_j(t))$ is a charge on $j$-th capacitor at instant $t$ [the voltage on this capacitor is equal to $V_j=V_j(t)$], while $I_j=I_j(t)$ represents current through the inductor $L_j$ \footnote{Note, that more realistic models can be found in~\cite{English}.}.

Taking into account relations $I_j=\dot{Q}_j$ and $Q_j(t)=\int^{V_j}_0 C(V)dV=C_0V_j(1-\frac{b}{3}V_j^2)$, one can reduce the two first-order Eqs.\eqref{kir} to the system of differential equations of the second order with respect to voltages $V_j(t)$:

\begin{equation}\label{model}
\ddot{V_j}(1-bV_j^2)-2bV_j(\dot{V_j})^2 = \gamma(V_{j-1} -2V_j+V_{j+1}).
\end{equation}
Hereafter all inductors are supposed to be identical: $L_j=L_0~(j=1..N$) and  $\gamma=1/(L_0C_0)$.
The appropriate scaling of variables in Eq.\eqref{model} allows one to suppose $b=1$ and $\gamma=1$. Finally, taking into account periodic boundary conditions, we obtain the following dynamical model (LC-chain) which is studied in the first part of the present paper (Secs. 2-9):

\begin{subequations}\label{model_with_bound}
   \begin{eqnarray}
   &\ddot{V_j}(1-V_j^2)-2V_j(\dot{V_j})^2 = (V_{j-1} -2V_j+V_{j+1}),\\
   &V_0(t) \equiv V_N(t), V_{N+1}(t) \equiv V_1(t).
   \end{eqnarray}
\end{subequations}

There is no resistance and, therefore, no energy dissipation in the LC-chains~\eqref{kir}. In Sec. 10, we consider a more general model which will be refered to as LCR-chain. It contains resistors ($R_J$) and a certan distribution of the external sources of time-periodic voltage.

The so-called zone boundary mode, or $\pi$-mode, which was investigated in Ref.~\cite{Osting}, can be written in the form
\begin{equation}\label{pi-anzats}
\boldsymbol{\phi}_1 =\{V(t), -V(t) |V(t), -V(t) |V(t),-V(t)| ...\}.
\end{equation}
This means that voltages on every pair of neighboring capacitors are opposite in sign. Obviously, such dynamical regime can exist only in chains with an even number of cells.

There are many papers devoted to study $\pi$-mode in monoatomic chains of the Fermi-Pasta-Ulam (FPU) type~\cite{zh-PhRevE,Bud-Bou,Page,Poggi,Yoshimura,FPU-1,FPU-2}. However, the model of LC-chain~\eqref{model_with_bound} essentially differs from the standard FPU-equations. In Appendix of Ref.~\cite{Osting}, the authors using some approximation reduce equations of the considered electric chain to those of the FPU-beta model, but we prefer to study the problem directly in the form~\eqref{model_with_bound}.

Stability of $\pi$-mode depends on the number of chain cells ($N$). Since this mode represents periodic regime, its stability can be investigated with the aid of the standard Floquet method. However, dimension of the variational system (the system of the equations linearized in the vicinity of the considered dynamical regime) and corresponding monodromic matrix is equal to $2N$. Because of this reason, studying of the linear stability of $\pi$-mode can be very difficult in the case $N\gg1$, especially when $N\rightarrow\infty$. A special method for studying this problem was developed in ~\cite{Osting}. However, this method is based essentially on the \emph{specific structure} of the dynamical model of the LC-chain, and seems to be rather difficult.

On the other hand, a general group-theoretical method for splitting (decomposing) variational system into independent subsystems, whose dimensions can be considerably smaller than that of the original variational system, was developed for dynamical models with discrete symmetry in our papers ~\cite{zh-PhRevE,zh-PND}. This method is based only on symmetry-related arguments and it uses the apparatus of irreducible representations of the symmetry group of the considered dynamical regime.

In Refs. ~\cite{FPU-1,FPU-2}, we used this method for analyzing linear stability of all possible symmetry-determined Rosenberg nonlinear normal modes in the FPU-$\alpha$ and FPU-$\beta$ chains. In Refs. ~\cite{2D-breathers,2D-breathers-a}, it was applied for studying stability of discrete breathers and quasibreathers in $2D$ scalar dynamical models on the plane square lattices.

In the present paper, we use the above group-theoretical method for analyzing stability of the symmetry-determined nonlinear normal modes in the electrical chains of LC and LCR types.

The paper is organized as follows. The possible symmetry-determined nonlinear normal modes (NNMs) in the LC-model~\eqref{model_with_bound} are discussed in Sect.~\ref{sec:2}. In Sect.~\ref{sec:3}, we briefly review the group-theoretical method for studying their stability. In Sect.~\ref{sec:4}, we apply this method for splitting variational systems for all NNMs permissible in the above model. Sect.~\ref{sec:5} is devoted to stability analysis of $\pi$-mode, while stability of others NNMs are discussed for the LC-chains in Sects.~\ref{sec:6}-\ref{sec:9}. In Sect.~\ref{sec:10}, we study the nonlinear normal modes and their stability in the LCR-chains. In Conclusion, we summarize the results on stability of NNMs in the considered electrical chain and discuss perspectives of applying our group-theoretical method for studying stability of quasiperiodic nonlinear vibrations (bushes of NNMs).

\section{Symmetry-determined nonlinear normal modes}\label{sec:2}

The concept of nonlinear normal modes was developed by Rosenberg in~\cite{rozenberg} (more information on these dynamical objects can be found in~\cite{vakakis}). In the dynamical regime corresponding to a given NNM, vibrations of all dynamical variables are described by one and the same time-dependent function $f(t)$. For the electrical model~\eqref{model_with_bound}, this means that all voltages $V_j(t)$ satisfies the relation
\begin{equation}\label{roz}
V_j(t)=c_jf(t), j=1..N,
\end{equation}
where $c_j$ are constant coefficients.

Note that convenient linear normal modes also satisfy the definition~\eqref{roz} with $f(t)=\sin(\omega t+\varphi_0)$, where $\omega$ and $\varphi_0$ represent frequency and initial phase, respectively.

Let us emphasize that Rosenberg nonlinear normal modes can exist only in some very specific classes of systems, for example, in systems whose potential energy is a homogeneous function of all its arguments. However, existence of NNMs in systems with general type of interactions can be ensured by the presence of some group of discrete symmetry~\cite{FPU-1,FPU-2,rink,CSSSH,PHD-1998}. Hereafter, we call such modes symmetry-determined nonlinear normal modes, or simply NNMs because only modes of this type are considered in the present paper.

It is very essential, that there exists only a small number of symmetry-determined NNMs in any lattice model. We discussed the cause of such situation in~\cite{FPU-2,CSSSH}. The matter is that the attempt of construction NNM with large cell in vibrational state leads to appearance of quasiperiodic dynamical regimes which represent bushes of NNMs ~\cite{DAN-1,DAN-2,PHD-1998} (see also ~\cite{Columbus,bountis,CSSSH}).

Let us give a brief review on this subject.

All possible dynamical regimes in physical system described by discrete symmetry group $G_0$ (the dynamical equations of the system are invariant according to all transformations of $G_0$) can be classified by \emph{subgroups} of the group $G_0$.

It is easy to check that equations~\eqref{model_with_bound} are invariant under the action of the group $G_0$ which is isomorphic to the symmetry group $D_n$ of the FPU-$\beta$ chain. This group can be described by generators which we define by their action on the vector
\begin{equation*}
	\textbf{V} = \{V_1(t), V_2(t), ... , V_N(t)\}
\end{equation*}
determining a given vibrational state of the electrical chain.

Transitional transformation $\hat{a}$ induces the cycle transposition of variables $V_j$ (note that periodic boundary conditions are assumed):
\begin{equation}\label{aV}
	\hat{a}\textbf{V} = \{V_2(t), V_3(t), ... V_N(t), V_1(t)\}.
\end{equation}
Transformation $\hat{\imath}$ generates inversion of the components of the vector $\textbf{V}$ with respect to an arbitrary chosen "middle" of the chain:

\begin{equation*}
	\hat{\imath}\textbf{V} = \{V_N(t), V_{N-1}(t),...,V_2(t), V_1(t)\},
\end{equation*}
Finally, the transformation $\hat{u}$ changes signs of all variables $V_j(t)$ without their transposition:
\begin{equation*}
	\hat{u}\textbf{V} = \{-V_1(t), -V_2(t), ..., -V_N(t)\}.
\end{equation*}

It is easy to check that the action of all generators of the group $G_0$, i.e., $\hat{a}, \hat{\imath},\hat{u}$, transforms the dynamical system~\eqref{model_with_bound} into itself and, therefore, they actually are invariant transformations for the considered electrical chain.

The orders of the generators $\hat{a},\hat{\imath},\hat{u}$ are equal to $N,2,2$ respectively, i.e.
\begin{equation}\label{orders}
\hat{a}^N=\hat{e}, \hat{\imath}^2=\hat{e}, \hat{u}^2=\hat{e},
\end{equation}
where $\hat{e}$ is the identical element of the group $G_0$. Thus, $G_0$ consist of $4N$ symmetry elements representing all products of the above three generators. Note that $G_0$ is a nonabelian group (for example, $\hat{\imath}\hat{a}\neq \hat{a}\hat{\imath}$).

Every subgroup $G_j~(j\neq 0)$ of the group $G_0$ determines a specific dynamical regime that can take place in the model~\eqref{model_with_bound}. The subgroup $G_j$ we call the symmetry group of the corresponding regime since the vector $\textbf{V}(t)$, describing this regime, is invariant under the action of all elements of $G_j$.

Let us consider the subgroup $G_j  \subset G_0$ that consists of identical element $\hat{e}$ and the transformation $\hat{a}^2$ for the electrical chain with $N=6$. We look for the dynamical regime $\textbf{V}(t)$ which is invariant under the action of this subgroup:
\begin{equation}\label{aV}
\hat{a}^2\textbf{V}=\textbf{V}.
\end{equation}
Taking into account that
\begin{equation}\label{90}
\hat{a}^2\boldsymbol{V}=(V_3, V_4, V_5, V_6, V_1, V_2),
\end{equation}
we obtain the following relations:
\begin{equation}\label{sootn}
V_3=V_1,\; V_4=V_2,\; V_5=V_3,\; V_6=V_4,\; V_1=V_5,\; V_2=V_6.
\end{equation}
Therefore, the six-dimensional vector $\textbf{V}$ is determined by \emph{only two} independent variables $V_1(t)$ and $V_2(t)$:
\begin{equation}\label{ab}
\textbf{V}=\{V_1,V_2|V_1,V_2|V_1,V_2\}.
\end{equation}

With the aid of relations~\eqref{sootn} between dynamical variables $V_j(t) (j=1..6)$ we reduce dynamical equations~\eqref{model_with_bound} to two independent equations with respect to variables $V_1(t)$ and $V_2(t)$ (all other equations~\eqref{model_with_bound} turn out to be equivalent to them). Thus, we obtain a \emph{two-parametric quasiperiodic} dynamical regime~\eqref{ab}: there are two basic frequencies and their integer combinations in the Fourier spectrum of such type of vibrations.

However, only periodic vibrations are considered in the present paper and, therefore, we must select from all symmetry-determined dynamical regimes only those which are \emph{one-parametric} (only one basic frequency presents in their Fourier spectrum). Obviously, these regimes represent Rosenberg nonlinear normal modes.

The problem of finding all one-parametric symmetry-determined dynamical regimes in a monoatomic chain can be solved with the aid of the group-theoretic method that was described in details in ~\cite{FPU-2}. (Note that the similar problem for all periodic structures described by any of 230 space groups was solved in ~\cite{CSSSH}).

Below, we present only final results of the selection of all symmetry-determined NNMs that are possible in chains~\eqref{model_with_bound} for different number $N$ of electrical cells.

It was found that only five types of such NNMs can exist in the considered LC-chains. Let us emphasize once more that this result does not depends on the specific structure of Eqs.~\eqref{model_with_bound} --- it is a consequence only of symmetry group of the considered lattice model.

Below, we list all symmetry-determined NNM for the model~\eqref{model_with_bound}.
\begin{align}
&G_1=[\hat{a}^2, \hat{\imath}\hat{u}]:\boldsymbol{\phi}_1 = V(t) \{ 1,-1|1,-1|1,-1|...|1,-1 \}\text{($\pi$-mode).}\label{pi_anzats}\\
&G_2=[\hat{a}^4, \hat{\imath}\hat{u}]:\boldsymbol{\phi}_2=V(t)\{1,1,-1,-1|1,1,-1,-1|...\}.\label{a2u-anzats}\\
&G_3=[\hat{a}^4, \hat{a}\hat{\imath}]:\boldsymbol{\phi}_3=V(t)\{1,0,-1,0|1,0,-1,0|...\}.\label{a4-anzats}\\
&G_4=[\hat{a}^3, \hat{\imath}\hat{u}]:\boldsymbol{\phi}_4=V(t)\{1,0,-1|1,0,-1|...\}.\label{a3-anzats}\\
&G_5=[\hat{a}^3\hat{u}, \hat{a}\hat{\imath}\hat{u}]:\boldsymbol{\phi}_5=V(t)\{1,1,0,-1,-1,0|1,1,0,-1,-1,0|...\}.\label{a6-anzats}
\end{align}

Let us comment on the notations used in Eqs.~\eqref{pi_anzats}-\eqref{a6-anzats}. Each subgroup $C_j\subset G_0 (j=1..5)$ is determined by two generators which are shown in square brackets. (Let us remind that the original group $G_0$ is determined by three generators: $G_0=[\hat{a},\hat{\imath},\hat{u}]$). For example, the generators $\hat{a}^2$ and $\hat{\imath}\hat{u}$ correspond to the subgroup $G_1=[\hat{a}^2,\hat{\imath}\hat{u}]$. The action of $\hat{a}^2$ on the vector $\textbf{V}(t)$ for the chain with $N=6$ was shown in Eq.~\eqref{90}, while the second generator, $\hat{\imath}\hat{u}$ acts on this vector as follows:
\begin{equation}\label{iuV}
	\hat{\imath}\hat{u}\textbf{V} = \{-V_6(t),-V_5(t),-V_4(t),-V_3(t), -V_2(t), -V_1(t)\}.
\end{equation}
Demanding invariance of the vector $\boldsymbol{V}(t)$ under the action of $\hat{a}^2$, we have obtained dynamical regime~\eqref{ab}. The presence of the second generator ($\hat{\imath}\hat{u}$) in the subgroup $G_1$ leads to an additional relation between variables $V_1(t), V_2(t)$:
\begin{equation}
V_2(t)=-V_1(t).
\end{equation}
Therefore, we obtain the one-dimensional regime representing $\pi$-mode~\eqref{pi_anzats}. Note that in Eqs.\eqref{pi_anzats}-\eqref{a6-anzats} we separate by vertical lines the different cells of the \emph{vibrational state} of the electrical chain. In both cases,~\eqref{ab} and~\eqref{pi_anzats}, this cell is twice larger than that of equilibrium state and includes two capacitors.

Vibrational cells including four neighbor capacitors correspond to the modes~\eqref{a2u-anzats} and~\eqref{a4-anzats}, and those embrasing three and six capacitors are associated with the modes~\eqref{a3-anzats} and~\eqref{a6-anzats}, respectively.

It is very essential that the modes with larger vibrational cells cannot exist in the considered electrical chain. Indeed, if we try to construct a dynamical regime of such type, it turns out to be quasiperiodic because of dependence of more than one independent variable (it represents a \emph{bush} of nonlinear normal modes with dimension $m>1$).

Substituting Eqs.\eqref{pi_anzats}-\eqref{a6-anzats} into dynamical equations~\eqref{model_with_bound}, we obtain only one independent equation with respect to the variable V(t):
\begin{equation}\label{ved}
  \ddot{V}(t)[1-V^2(t)]-2V(t)[\dot{V}(t)]^2+\mu V(t)=0,
\end{equation}
where the following values of the parameter $\mu$ correspond to the NNMs shown in Eqs.~\eqref{pi_anzats}-\eqref{a6-anzats}:

$G_1=[\hat{a}^2, \hat{\imath}\hat{u}]:\mu=4,$

$G_2=[\hat{a}^4, \hat{\imath}\hat{u}]:\mu=2,$

$G_3=[\hat{a}^4, \hat{a}\hat{\imath}]:\mu=2,$

$G_4=[\hat{a}^3, \hat{\imath}\hat{u}]:\mu=3,$

$G_5=[\hat{a}^3\hat{u}, \hat{a}\hat{\imath}\hat{u}]:\mu=1.$

Hereafter, we call  Eq.~\eqref{ved} \emph{governing} equation for the corresponding nonlinear normal mode.

\section{The method for studying nonlinear normal mode stability}\label{sec:3}

A given NNM represents periodic dynamical regime and, therefore, the standard Floquet method can be used for studying its stability. Using this method, we linearize the original system of nonlinear differential equations near this NNM and, as a result, we obtain the so called \emph{variational} system. This is a system of $N$ second-order differential equations with time-periodic coefficients, whose period $T$ is equal to that of the given NNM. Then we reduce this system to the system $F$ of $2N$ first-order differential equations and construct for it the \emph{monodromic} matrix. The successive \emph{columns} of this matrix can be obtained by integrating $2N$ times the system $F$ over period $T$ using the columns of $2N\times 2N$ identical matrix as the corresponding initial conditions.

As was already noted, the above-mentioned Floquet procedure is very difficult for the case $N\gg 1$, especially when $N\rightarrow \infty$. For the dynamical systems with discrete symmetries these difficulties can be often overcome with the aid of the group-theoretical method developed in our papers~\cite{zh-PhRevE,zh-PND}. Below, we give brief outline of this method.

Let us consider the symmetry group $G_0$ of the dynamical model~\eqref{model_with_bound}. Every element $g\in G_0$ induces a certain transposition of the variables $V_j~(j=1..N)$, possibly, accompanied by changing signs of some of these variables. Noting that $\ddot{V}_j$ and $\dot{V}_j$ are transformed under the action of the element $g$ as variables $V_j$, we conclude that this element leads to the same transposition of equations~\eqref{model_with_bound} with the same changing of their signs as the set $V_j~(j=1..N)$.

Obviously, arbitrary chosen transposition of variables $V_j$ may not belong to the group $G_0$. For example, if we transpose only two variables, $V_2\leftrightarrow V_3$, without any transposition of other variables, equations~\eqref{model_with_bound} change essentially. Indeed, equations 2 and 3 of the system~\eqref{model_with_bound} read
\begin{equation*}
\begin{split}
\ddot{V_2}(1-V_2^2)-2V_2(\dot{V_2})^2 = (V_1 -2V_2+V_3),\\
\ddot{V_3}(1-V_3^2)-2V_3(\dot{V_3})^2 = (V_2 -2V_3+V_4).
\end{split}
\end{equation*}
After the above transposition these equations acquire the following form:
\begin{equation*}
\begin{split}
\ddot{V_3}(1-V_3^2)-2V_3(\dot{V_3})^2 = (V_1 -2V_3+V_2),\\
\ddot{V_2}(1-V_2^2)-2V_2(\dot{V_3})^2 = (V_3 -2V_2+V_4).
\end{split}
\end{equation*}
Since equations 2 and 3 don't transform into each other as a result of the transposition $V_2 \leftrightarrow V_3$ [obviously, their coincidence with any other equations~\eqref{model_with_bound} is excluded], this transformation cannot belong to the group $G_0$.

In this paper, we don't discuss any methods for revealing the symmetry group $G_0$ of the given dynamical model~\eqref{model_with_bound} (one such method was developed in ~\cite{FPU-2}). Note that all symmetry elements of the system equilibrium state must enter the group $G_0$.

Let us consider a dynamical regime in our electrical chain~\eqref{model_with_bound}, described by the vector $\textbf{V}(t)=\{V_1(t),V_2(t),..V_n(t)\}$. Acting on this vector by all elements $g\in G_0$, we choose only those which don't change it, i.e. elements satisfying condition $\hat{g}\textbf{V}=\textbf{V}$. Full set of such elements represents a subgroup $G$ of the group $G_0$: $G\subset G_0$. This subgroup is the \emph{symmetry group} of the considered dynamical regime $\textbf{V}(t)$.

According to the theorem proved in ~\cite{zh-PhRevE}, the variational system obtained by linearization of nonlinear equations~\eqref{model_with_bound} in the vicinity of the dynamical regime $V(t)$ is invariant with respect to the above subgroup $G\subset G_0$. This means that some elements $g\in G_0$ which conserve the original system~\eqref{model_with_bound} don't conserve the corresponding variational system.

The action of a symmetry element $g\in G$ on the vector $\textbf{V}$ can be replaced by the action of a certain $N\times N$ square matrix. The set of such matrices associated with all elements $g\in G$, acting in $N$-dimensional space of state vectors $\boldsymbol{V}$, represents a matrix representation $\Gamma$ of the group $G$.

The condition that $G$ is the symmetry group of the variational system $\ddot{\boldsymbol{\delta}}=J(t)\boldsymbol{\delta}$ (this is the form of the variational system for the FPU-$\beta$ model) leads to commutation of the matrix $J(t)$ with all matrices of the representation $\Gamma$ ~\cite{zh-PhRevE}. This fact allows one to apply the well-known Wigner theorem describing the structure of the matrix commuting with all matrices of a \emph{reducible} representation of a given finite group.

The conclusion of the Wigner theorem can be explain as follows. Let us decompose the reducible representation $\Gamma$ of the group $G$ into irreducible representations (irreps) $\Gamma_j$ of the same group:
\begin{equation}\label{sum_rep}
\Gamma={\sum_j}^\oplus m_j\Gamma_j
\end{equation}
($m_j$ are nonnegative integers).
Such decomposition can be fulfilled by the appropriate transformation of the basis of $N$-dimensional vector space corresponding to $\Gamma$ (carrier space). Then matrix $J(t)$, commuting with all matrices of $\Gamma$, acquires a block-diagonal form as a result of the above-mentioned transformation of the original basis. Every block $D_j$ corresponds to a certain irrep $\Gamma_j$ and possesses dimension equal to $n_j m_j$, where $n_j$ is the dimension of $\Gamma_j$, while $m_j$ [see Eq.~\eqref{sum_rep}] determines how many times $\Gamma_j$ enters the reducible representation $\Gamma$. Moreover, the block $D_j$ represents a matrix with very specific structure: it is a \emph{direct product} of a certain $m_j$-dimensional matrix and $n_j$-dimensional identical matrix.

As was shown in ~\cite{zh-PhRevE}, this structure of the matrix $J=J(t)$, obtained after basis transformation producing the form~\eqref{sum_rep} of the representation $\Gamma$, leads to \emph{splitting} (decomposition) of the original variational system $\ddot{\boldsymbol{\delta}}=J(t)\boldsymbol{\delta}$ into a number of \emph{independent} subsystems whose dimensions can be considerably smaller than that of the original variational system. Namely, $n_j$ identical subsystems of dimension $m_j$ correspond to every block $D_j$ and, therefore, to the irrep $\Gamma_j$. Note, that induction numbers $m_j$ entering Eq.~\eqref{sum_rep} can be found with the aid of the following formula
\begin{equation}\label{kratn}
	m_i=\frac{1}{\|G\|}\sum_{g\in G}  {\chi}_{\scriptscriptstyle\Gamma}(g)\bar{\chi}_{\scriptscriptstyle\Gamma_i}(g)
\end{equation}
from the theory of finite groups~\cite{Petrashen}. Here, $\|G\|$ is the order of the group $G$ (the number of its elements), while $\chi_{\scriptscriptstyle\Gamma}(g)$ and $\bar{\chi}_{\scriptscriptstyle\Gamma_i}(g)$ are traces of matrices corresponding to the element $g$ in the representations $\Gamma$ and $\Gamma_j$, respectively. The bar over $\bar{\chi}_{\scriptscriptstyle\Gamma_i}(g)$ denotes complex conjugation.

One can obtain the \emph{explicit} form of variational system $\ddot{\boldsymbol{\delta}}=J(t)\boldsymbol{\delta}$ decomposed into independent subsystems by passing from the original basis in the space of all infinitesimal vectors $\boldsymbol{\delta}$ to the new one which represents the set $W$ of all basis vectors of the irreps $\Gamma_j$ entering the decomposition~\eqref{sum_rep}. This transformation of the basis leads to transformation of the old infinitesimal variables $\delta_i(t)$ to the new one $y_i(t)$ as follows:
\begin{equation}\label{y300}
\boldsymbol{y}=S\boldsymbol{\delta},
\end{equation}
where the columns of the orthogonal matrix $S$ are vectors from $W$.

Let us illustrate the above described method with a simple example.

For the case $N=4$, the variational system of the electrical chain~\eqref{model_with_bound} reads:

\begin{equation}\label{var_sys_4}
\left\{
\begin{aligned}
   &(1-V_1^2)\ddot{\delta_1}-2\dot{(V_1^2)}\dot{\delta_1}=2(\ddot{V_1\dot{V_1}}-1)\delta_1+\delta_2+\delta_4,\\
   &(1-V_2^2)\ddot{\delta_2}-2\dot{(V_2^2)}\dot{\delta_2}=2(\ddot{V_2\dot{V_2}}-1)\delta_2+\delta_3+\delta_1,\\
   &(1-V_3^2)\ddot{\delta_3}-2\dot{(V_3^2)}\dot{\delta_3}=2(\ddot{V_3\dot{V_3}}-1)\delta_3+\delta_4+\delta_2,\\
   &(1-V_4^2)\ddot{\delta_4}-2\dot{(V_4^2)}\dot{\delta_4}=2(\ddot{V_4\dot{V_4}}-1)\delta_4+\delta_1+\delta_3.\\
\end{aligned}
\right.
\end{equation}
Here $V_i=V_i(t), \delta_i=\delta_i(t), i=1..4$.

Let $\alpha_i=1-V_i^2, \beta_i=-2\dot{(V_i^2)},\gamma_i=2(\ddot{V}_i\dot{V}_i-1)$. Then Eqs.~\eqref{var_sys_4} can be written in the matrix form:
\begin{equation}\label{var_sys_mat}
\hat{\alpha}(t)\ddot{\boldsymbol{\delta}}+\hat{\beta}(t)\dot{\boldsymbol{\delta}}=\hat{\gamma}(t)\boldsymbol{\delta},
\end{equation}
where matrices $\hat{\alpha}(t)$, $\hat{\beta}(t)$ and $\hat{\gamma}(t)$ read
\begin{equation}
\hat{\alpha}(t) = \begin{pmatrix} \alpha_1 & 0 & 0 & 0 \\ 0 & \alpha_2 & 0 & 0 \\ 0 & 0 & \alpha_3 & 0 \\ 0 & 0 & 0 & \alpha_4 \end{pmatrix},
\hat{\beta}(t) = \begin{pmatrix} \beta_1 & 0 & 0 & 0 \\ 0 & \beta_2 & 0 & 0 \\ 0 & 0 & \beta_3 & 0 \\ 0 & 0 & 0 & \beta_4 \end{pmatrix},
\end{equation}
\begin{eqnarray}
&\hat{\gamma}(t) =\hat{\gamma}_1(t)+\hat{\gamma}_2,\\
&\hat{\gamma}_1(t)=\begin{pmatrix} \gamma_1 & 0 & 0 & 0 \\ 0 & \gamma_2 & 0 & 0 \\ 0 & 0 & \gamma_3 & 0 \\ 0 & 0 & 0 & \gamma_4 \end{pmatrix}, \hat{\gamma}_2=\begin{pmatrix} 0 & 1 & 0 & 1 \\ 1 & 0 & 1 & 0 \\ 0 & 1 & 0 & 1 \\ 1 & 0 & 1 & 0 \end{pmatrix}.
\end{eqnarray}

One possible dynamical regime in the model~\eqref{model_with_bound} is
\begin{equation}\label{v1v2}
\textbf{V}(t)=\{U_1(t),U_2(t),U_2(t),U_1(t)\}.
\end{equation}
This fact can be checked by straightforward substitution of this vector into Eqs.~\eqref{model_with_bound}. The four equations with respect to variables $V_i(t),i=1..4$ reduce to two equations for new variables $U_1(t)$ and $U_2(t)$.

Note that the vector~\eqref{v1v2} determines a \emph{quasiperiodic} dynamical regime, and we want to emphasize once more that our method can be applied not only to periodic, but also to quasiperiodic regimes.

It is easy to find the symmetry group $G$ of the dynamical regime~\eqref{v1v2}. This is a second-order group
\begin{equation}\label{v1v2group}
G=[\hat{e}, \hat{\imath}]
\end{equation}
whose nontrivial element $\hat{\imath}$ acts on the arbitrary state vector $\textbf{V}$ as follows:
\begin{equation}
\hat{\imath}\{V_1(t),V_2(t),V_3(t),V_4(t)\}=\{V_4(t),V_3(t),V_2(t),V_1(t)\}.
\end{equation}
It transposes voltages as $V_2\leftrightarrow V_3,V_1\leftrightarrow V_4$ and turns out to be a certain symmetry transformation of the model~\eqref{model_with_bound}. Obviously, this transformation is simultaneously the symmetry element of the regime~\eqref{v1v2}.

As any second-order group, the group~\eqref{v1v2group} possesses two irreducible representations, $\Gamma_1$ and $\Gamma_2$, both one-dimensional, which we present in Table~\ref{tablt}.

\begin{table}[h]\label{tablt}
\caption{Irreducible representations of the group $G=[\hat{e},\hat{\imath}]$}
\label{tablt}
\begin{center}
\begin{tabular}{|p{6em}|p{6em}|p{6em}|}
  \hline
   Irreps & $\hat{e}$ & $\hat{\imath}$ \\
  \hline
  $\Gamma_1$ & 1 & 1 \\
  $\Gamma_2$ & 1 & -1 \\
  \hline
\end{tabular}
\end{center}
\end{table}

Now we must construct basis vectors, $\boldsymbol{\varphi}$ and $\boldsymbol{\psi}$, of these irreps, which represent certain four-dimensional vectors. Let us remember the definition of the representation $\Gamma$ (reducible or irreducible) of an arbitrary finite group $G$:
\begin{equation}\label{rep}
\hat{g}\boldsymbol{\phi}=\tilde{M}(g)\boldsymbol{\phi},\ \forall g \in G
\end{equation}
Here $\boldsymbol{\phi}=\{\boldsymbol{\phi}_1,\boldsymbol{\phi}_2,...,\boldsymbol{\phi}_N\}$ is the "supervector", representing the list of basis vectors of a given $n$-dimensional representation, $M(g)$ is the matrix associated in this representation with element $g\in G$, while $\tilde{M}(g)$ denotes the transpose of the matrix $M(g)$. Eq.~\eqref{rep} represents a system of linear algebraic equations for unknown vectors $\boldsymbol{\phi}_i, i=1..n$.

Traditionally, basis vectors of an irreducible representation are found by means of projection operators (see, for example~\cite{Petrashen}). However, it is more convenient for our purpose to obtain these vectors using the straightforward method \cite{40} based on the definition~\eqref{rep} of matrix representation (some details of this method can be found in~\cite{zh-PhRevE}). Obviously, it is sufficient to use equations~\eqref{rep} only for the generators of the group $G$.

Let us find the basis vectors $\boldsymbol{\varphi}$ and $\boldsymbol{\psi}$ of the irreps $\Gamma_1$ and $\Gamma_2$, whose one-dimensional matrices are given in Table 1.

From Eqs.~\eqref{rep} we obtain for $\Gamma_1$ and $\Gamma_2$ the following equations
\begin{equation}\label{repg1}
\Gamma_1:\hat{\imath}\boldsymbol{\varphi}=(1)\boldsymbol{\varphi},
\end{equation}
\begin{equation}\label{repg2}
\Gamma_2:\hat{\imath}\boldsymbol{\psi}=(-1)\boldsymbol{\psi}.
\end{equation}
We search $\boldsymbol{\varphi}$ and $\boldsymbol{\psi}$ in the form $\boldsymbol{\varphi}=(V_1,V_2,V_3,V_4)$ and $(U_1,U_2,U_3,U_4)$, where $V_i,U_i(i=1..4)$ are arbitrary scalar values. From Eq.~\eqref{repg1}, we find
\begin{equation*}
\hat{\imath}\boldsymbol{\varphi}=\hat{\imath}(V_1,V_2,V_3,V_4)=(V_4,V_3,V_2,V_1)=(V_1,V_2,V_3,V_4).
\end{equation*}
Therefore, $V_1=V_4,V_2=V_3$ and
\begin{equation}\label{phig1}
\boldsymbol{\varphi}=(V_1,V_2,V_2,V_1).
\end{equation}
From Eq.~\eqref{repg2}, we obtain
\begin{equation*}
\hat{\imath}\boldsymbol{\psi}=\hat{\imath}(U_1,U_2,U_3,U_4)=(U_4,U_3,U_2,U_1)=(-U_1,-U_2,-U_3,-U_4).
\end{equation*}
Therefore, $U_4=-U_1,U_3=-U_2$ and
\begin{equation}\label{psig2}
\boldsymbol{\psi}=(U_1,U_2,-U_2,-U_1).
\end{equation}
In our example, both vectors $\boldsymbol{\psi}$ and $\boldsymbol{\varphi}$ represent two-dimensional subspaces of the space of all state vectors, and they can be written in the form:
\begin{equation}\label{phi-vect}
\boldsymbol{\varphi}=V_1
\begin{pmatrix} 1\\ 0\\ 0\\ 1\end{pmatrix}
+V_2\begin{pmatrix} 0\\ 1\\ 1\\ 0\end{pmatrix},
\end{equation}
\begin{equation}\label{psi-vect}
\boldsymbol{\psi}=V_1
\begin{pmatrix} 1\\ 0\\ 0\\ -1\end{pmatrix}
+V_2\begin{pmatrix} 0\\ 1\\ -1\\ 0\end{pmatrix}.
\end{equation}
Each of these vectors determines two basis vectors of the four-dimensional space of the nonlinear model~\eqref{model_with_bound} or linear model~\eqref{var_sys_4} for $N=4$. Normalizing these basis vectors, we can write them as a columns of the matrix $S$ entering Eq.~\eqref{y300}:
\begin{equation}\label{matrixS}
S=\frac{1}{\sqrt{2}}\begin{pmatrix} 1 & 0 & 1 & 0 \\ 0 & 1 & 0 & 1 \\ 0 & 1 & 0 & -1 \\ 1 & 0 & -1 & 0 \end{pmatrix}.
\end{equation}
The columns of the matrix $\hat{S}$ represent the orthogonal and normalized basis in the space of all variables of the dynamical models~\eqref{model_with_bound} and~\eqref{var_sys_4} for the case $N=4$.

Using the matrix~\eqref{matrixS}, we can split the variational system in the form~\eqref{var_sys_4} or~\eqref{var_sys_mat} into two independent subsystems.

Indeed, for dynamical regime~\eqref{v1v2}, matrices $\hat{\alpha},\hat{\beta}, \hat{\gamma}$ acquire the forms
\begin{equation}
\hat{\alpha}(t) = \begin{pmatrix} \alpha_1 & 0 & 0 & 0 \\ 0 & \alpha_2 & 0 & 0 \\ 0 & 0 & \alpha_2 & 0 \\ 0 & 0 & 0 & \alpha_1 \end{pmatrix},
\end{equation}
\begin{equation}
\hat{\beta}(t) = \begin{pmatrix} \beta_1 & 0 & 0 & 0 \\ 0 & \beta_2 & 0 & 0 \\ 0 & 0 & \beta_2 & 0 \\ 0 & 0 & 0 & \beta_1 \end{pmatrix},
\end{equation}
\begin{equation}
\hat{\gamma}(t) = \begin{pmatrix} \gamma_1 & 1 & 0 & 1 \\ 1 & \gamma_2 & 1 & 0 \\ 0 & 1 & \gamma_2 & 1 \\ 1 & 0 & 1 & \gamma_1 \end{pmatrix}.
\end{equation}
After orthogonal transformation with the matrix $\hat{S}$ from Eq.~\eqref{matrixS}, we obtained:
\begin{equation}
\hat{\alpha}(t) = \begin{pmatrix} \alpha_1 & 0 & 0 & 0 \\ 0 & \alpha_2 & 0 & 0 \\ 0 & 0 & \alpha_1 & 0 \\ 0 & 0 & 0 & \alpha_2 \end{pmatrix},
\end{equation}
\begin{equation}
\hat{\beta}(t) = \begin{pmatrix} \beta_1 & 0 & 0 & 0 \\ 0 & \beta_2 & 0 & 0 \\ 0 & 0 & \beta_1 & 0 \\ 0 & 0 & 0 & \beta_2 \end{pmatrix},
\end{equation}
\begin{equation}
\hat{\gamma}(t) = \begin{pmatrix} \gamma_1+1 & 1 & 0 & 0 \\ 1 & \gamma_2+1 & 0 & 0 \\ 0 & 0 & \gamma_1-1 & 1 \\ 0 & 0 & 1 & \gamma_2-1 \end{pmatrix}.
\end{equation}
Each of the above matrices can be written as direct sums of certain two-dimensional blocks. Therefore, the variational system~\eqref{var_sys_4} is decomposed [in the variables $y_j~(j=1..4)$ from Eq.~\eqref{y300}] into two independent subsystems of the form:
\begin{subequations}
\begin{eqnarray}
&\begin{pmatrix} \alpha_1(t) & 0 \\ 0 & \alpha_2(t) \end{pmatrix}\ddot{\boldsymbol{y}}+\begin{pmatrix} \beta_1 (t) & 0 \\ 0 & \beta_2(t)\end{pmatrix}\dot{\boldsymbol{y}}=\begin{pmatrix} \gamma_1(t)+1 & 1 \\ 1 & \gamma_2(t)+1 \end{pmatrix}\boldsymbol{y},\\
&\begin{pmatrix} \alpha_1(t) & 0 \\ 0 & \alpha_2(t) \end{pmatrix}\ddot{\boldsymbol{y}}+\begin{pmatrix} \beta_1 (t) & 0 \\ 0 & \beta_2(t)\end{pmatrix}\dot{\boldsymbol{y}}=\begin{pmatrix} \gamma_1(t)-1 & 1 \\ 1 & \gamma_2(t)-1\end{pmatrix}\boldsymbol{y},
\end{eqnarray}
\end{subequations}
where $\boldsymbol{y}=(y_1, y_2)$.

Finally, let us note that as a result of the relations~\eqref{phi-vect} and~\eqref{psi-vect}, one can write the decomposition formula~\eqref{sum_rep} for our case as follows:
\begin{equation}
\Gamma=2\Gamma_1+2\Gamma_2.
\end{equation}

\section{Variational system splitting for the NNMs~\eqref{pi_anzats}-\eqref{a6-anzats}}\label{sec:4}

Each nonlinear normal mode from Eqs.~\eqref{pi_anzats}-\eqref{a6-anzats} possesses a certain symmetry group $G$ which contains a translational subgroup $T_n$. Let us consider the mode $\boldsymbol{\phi}_2$. It can be excited in chains for which $n=\frac{N}{4}$ is an integer number ($N~mod~ 4=0$). The mode $\boldsymbol{\phi}_2$ [see Eq.~\eqref{a2u-anzats}] is determined by extended primitive cell (EPS) of the form $\{V(t), V(t), -V(t),-V(t)\}$ which is repeated along the chain. This EPS being four times larger than the cell in equilibrium state describes a specific vibrational regime of voltages in the electrical chain~\eqref{model_with_bound} under the periodic boundary conditions.

Obviously, translation of the voltage distribution~\eqref{a2u-anzats} by $4a$, where $a$ is the cell size in equilibrium state, does not change  the form of this distribution and, therefore, $\hat{a}^4$ is one of its symmetry element. This element is a generator of the translational symmetry $T_n$.

The same translational group corresponds to the mode $\boldsymbol{\phi}_3$~\eqref{a4-anzats}. However, the full symmetry groups of the modes $\boldsymbol{\phi}_2$ and $\boldsymbol{\phi}_3$ are different: they differ by the second generator, $\hat{\imath}\hat{u}$ for $\boldsymbol{\phi}_2$ and $\hat{a}\hat{\imath}$ for $\boldsymbol{\phi}_3$.

One can split variational system for $\boldsymbol{\phi}_2$ and $\boldsymbol{\phi}_3$ by the full symmetry groups of the corresponding dynamical regimes, $[\hat{a}^4,\hat{\imath}\hat{u}]$ and $[\hat{a}^4,\hat{a}\hat{\imath}]$, but for our purpose it is more convenient to use a common group $G=[\hat{a}^2\hat{u}]$ of these modes which is defined by the generator $\hat{a}^2\hat{u}$. The action of this symmetry element on the voltage distribution $\{V_j(t)|j=1..N\}$ reduces to $2a$-translation accompanying by change of sign of all variables $V_j(t)$. Note that square of the symmetry element $\hat{a}^2\hat{u}$ is equal to the above discussed translational generator $\hat{a}^4$. Using the group $G=[ \hat{a}^2u]$, we can split the variational systems for both NNMs $\boldsymbol{\phi}_2$ and $\boldsymbol{\phi}_3$ with one and the same matrix $S$.

Now according to the general prescription, we must construct the basis vectors of the irreps of the group $G$. All such representations are one-dimensional, because this group is cyclic and, therefore, abelian. Every irrep $\Gamma_j$ of $G$ can be fully determined by the one-dimensional matrix $\gamma_j$ associated with the generator $\hat{a}^2\hat{u}$ of this group. The number of irreps of the group $G$ is equal to its order $n=\frac{N}{2}$. We can define these irreps, $\Gamma_j$, by $1\times1$ matrices: $\gamma_j=e^{\frac{2\pi i j}{n}}, j=0,1,2,...(n-1)$ [$\gamma_j$ are roots of unity order $n$].

As was described in the previous section, the basis vectors $\boldsymbol{\psi}_j$ of one-dimensional irreps $\Gamma_j$ can be obtained by solving the following equation
\begin{equation}\label{a2ueqval}
\hat{a}^2\hat{u}\boldsymbol{\psi}_j=\gamma_j\boldsymbol{\psi}_j,
\end{equation}
where $\boldsymbol{\psi}_j=\{V_1,V_2,..V_N\}$ is an arbitrary voltage distribution for our electrical chain. For the considered case, we can easily find (see some details in~\cite{zh-PhRevE}) that the solution to Eq.~\eqref{a2ueqval} represents a two-dimensional space $L$ of all possible voltage distributions which can be written in the form
\begin{equation}\label{psiab}
\begin{aligned}
\boldsymbol{\psi_j}=&\{|a,b,-\gamma_j^{-1}a, -\gamma_j^{-1}b|\gamma_j^{-2}a, \gamma_j^{-2}b,-\gamma_j^{-3}a, -\gamma_j^{-3}b|\\
&|\gamma_j^{-4}a, \gamma_j^{-4}b,-\gamma_j^{-5}a,-\gamma_j^{-5}b|...\},
\end{aligned}
\end{equation}
where $a$ and $b$ are arbitrary constants. Setting $a=1, b=0$ and $a=0, b=1$, we select two basis vectors of the above subspace.

Finally, looking over all values $j=0, 1, 2,...,(\frac{N}{2}-1)$, we find a certain basis of the full space $L$:
\begin{equation}\label{vectors_psi12}
\begin{aligned}
\boldsymbol{\psi}_j^{(1)}&=\{|1,0,-\gamma_j^{-1},0|\gamma_j^{-2},0,-\gamma_j^{-3},0|\gamma_j^{-4},0,-\gamma_j^{-5},0|...\},\\
\boldsymbol{\psi}_j^{(2)}&=\{|0,1,0,-\gamma_j^{-1}|0,\gamma_j^{-2},0,-\gamma_j^{-3},0|0,\gamma_j^{-4},0,-\gamma_j^{-5}|...\}.
\end{aligned}
\end{equation}

After normalization, we can use vectors~\eqref{vectors_psi12} as columns of the matrix $S$ which splits the variational systems for $\boldsymbol{\phi}_2$ and $\boldsymbol{\phi}_3$ into independent subsystems. Note that taking into account any additional symmetry elements beyond those of the group $G=[\hat{a}^2\hat{u}]$ can generally reduce dimensions of the above subsystems that were obtained with the aid of the group $G$.

Proceeding in such a manner, we reveal that the variational systems for studying stability of NNMs~\eqref{pi_anzats}-\eqref{a6-anzats} can be split into independent subsystems whose dimensions are equal to 1, 2, 2, 3, 3, respectively.

Below, we present these subsystems in matrix form
\begin{equation}\label{SplitMatrix}
\hat{A}\ddot{\boldsymbol{\delta}}+\hat{B}\dot{\boldsymbol{\delta}}+\hat{D}\boldsymbol{\delta}=0,
\end{equation}
where matrices $\hat{A}, \hat{B}, \hat{D}$ are constructed from the following elements:
\begin{equation}\label{cfg_def}
\begin{aligned}
&c(t)=1-V^2,\\
&f(t)=2(\dot{V}^2+V\ddot{V}-1),\\
&g(t)=-4V\dot{V}.\\
\end{aligned}
\end{equation}
Here $V=V(t)$ is the solution of the governing equation~\eqref{ved} for the corresponding nonlinear normal mode.

1) $\boldsymbol{\phi}_1=V(t)\{1,-1|1,-1|1,-1|...\}$.

For this case all subsystems~\eqref{SplitMatrix} are one-dimensional with the matrices
\begin{equation}\label{PiM}
\hat{A}=c(t), \hat{B}=g(t), \hat{D}=f(t)-2\cos(k),
\end{equation}
where $k=\frac{2\pi j}{N}$.

For NNMs $\boldsymbol{\phi}_2$ and $\boldsymbol{\phi}_3$, subsystems~\eqref{SplitMatrix} are two-dimensional:

2)$\boldsymbol{\phi}_2=V(t)\{1,1,-1,-1|1,1,-1,-1|...\}$.
\begin{equation}\label{a2uM}
\hat{A}=\begin{pmatrix} c & 0\\ 0 & c\end{pmatrix}, \hat{B}=\begin{pmatrix} g & 0\\ 0 & g\end{pmatrix},
\hat{D}=\begin{pmatrix} f & 1+\gamma\\ 1+\bar{\gamma}& f\end{pmatrix}.
\end{equation}

3)$\boldsymbol{\phi}_3=V(t)\{1,0,-1,0|1,0,-1,0|...\}$.
\begin{equation}\label{a4M}
\hat{A}=\begin{pmatrix} c & 0\\ 0 & 1\end{pmatrix}, \hat{B}=\begin{pmatrix} g & 0\\ 0 & 0\end{pmatrix},
\hat{D}=\begin{pmatrix} f & 1+\gamma\\ 1+\bar{\gamma}& -2\end{pmatrix}.
\end{equation}
In Eqs.~\eqref{a2uM}-\eqref{a4M}, $\gamma=e^{\frac{4\pi j}{N}j}$.

4) For both NNMs $\boldsymbol{\phi}_4=V(t)\{1,0,-1|1,0,-1|...\}$ and

$\boldsymbol{\phi}_5=V(t)\{1,1,0,-1,-1,0|1,1,0,-1,-1,0\}$
three-dimensional subsystems~\eqref{SplitMatrix} turn out to be \emph{identical}. These subsystems are determined by the matrices
\begin{equation}\label{a3M}
\hat{A}=\begin{pmatrix} c & 0 & 0\\ 0 & 1 & 0\\ 0 & 0 & c\end{pmatrix}, \hat{B}=\begin{pmatrix} g & 0 & 0\\ 0 & 0 & 0\\ 0& 0 & g\end{pmatrix},
\hat{D}=\begin{pmatrix} f & 1 & \gamma \\ 1 & -2 & 1 \\ \bar{\gamma} & 1 & f\end{pmatrix},
\end{equation}
where $\gamma=e^{\frac{6\pi i}{N}j}$.

Thus, stability of all NNMs in the LC-chain~\eqref{model_with_bound} can be investigated by analysing a set of one-, two- and three-dimensional systems of differential equations with time-periodic coefficients determined by the function $V(t)$.

\section{ Stability analysis of $\pi$-mode}\label{sec:5}

As we have just seen, the most simple splitting takes place for $\pi$-mode. Indeed, in this case the original variational system for the chain with any even\footnote{Note, that condition $N~mod~2=0$ is \emph{necessary} for existence of $\pi$-mode.} number of cells ($N$) can be split into one-dimensional subsystems which are determined by Eqs.~\eqref{SplitMatrix}-\eqref{PiM}. In the explicit form these equations read
\begin{equation}\label{Pi_explicit_form}
(1-V^2)\ddot{\delta}-4(V\dot{V})\dot{\delta}-2(\dot{V}^2+2V\ddot{V})\delta=-4\cos^2(\frac{k}{2})\delta,
\end{equation}
where $k=\frac{2\pi j}{N}, j=1,2,...,\frac{N}{2}$.

Eq.~\eqref{Pi_explicit_form} is a linear second-order differential equation with first derivative. One can reduce it to the so-called normal form which doesn't contain the first derivative by the appropriate changing of variables $\delta$ and $t$~\cite{kamke}. In our case, this procedure is simplifyed essentially, because the left part of Eq.~\eqref{Pi_explicit_form} may be written as the second derivative of the expression $[(1-V^2)\delta]$. Therefore, introducing the new infinitesimal variable
\begin{equation}\label{y-delta}
y(t)=[1-V^2(t)]\delta(t),
\end{equation}
we immediately obtain a very simple equation
\begin{equation}\label{pisimpl}
\ddot{y}+4c(t)\cos^2(\frac{k}{2})y=0,
\end{equation}
where $c(t)=[1-V^2(t)]^{-1}$.

Let us comment on the notation used in this equation. The time-periodic function
\begin{equation}\label{capacity}
c(t)=1/[1-V^2(t)]
\end{equation}
is determined by governing equation~\eqref{ved} with $\mu =4$ for initial conditions
 \begin{equation}\label{icsPi}
 V(0)=A, \dot{V}(t)=0.
 \end{equation}
Thus, $A$ is the amplitude of voltage oscillations on capacitors. The variable $k=\frac{2\pi}{N}j, ~j=1..N$ is introduced instead of the mode number $j$ (for the limit case $N \rightarrow \infty$ it represents the one-dimensional wave vector).

As was shown in Ref.~\cite{Osting}, the governing equation~\eqref{ved} describes a periodic dynamical regime for $A<1$. The value $A=1$ represents singular point beyond which the motion becomes infinite.

\begin{figure}[htb]
\begin{center}
\includegraphics[scale=0.8]{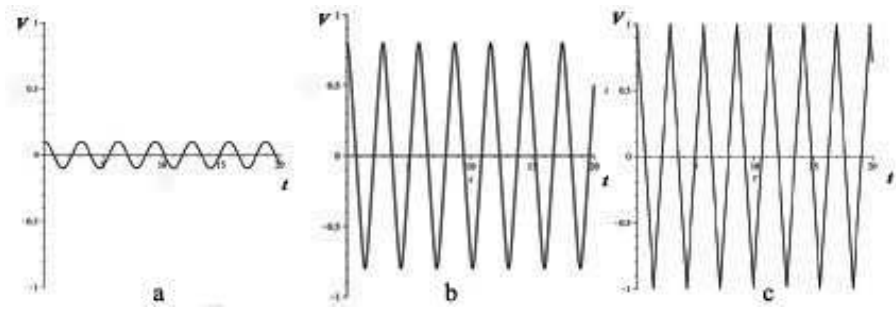}
\caption{Function $V(t)$ for three values of $\pi$-mode amplitude: a) $A=0.1$; b) $A=0.8$; c) $A=0.999$}{\label{ampl} }
\end{center}
\end{figure}

In Fig.\ref{ampl}, we present plots of function $V(t)$ for three values of its amplitude: $A=0.1, A=0.8, A=0.99$. As one can see, practically harmonic oscillations take place for small amplitude ($A=0.1$), while nonlinearity of vibrations is observed clearly for $A=0.8$. For $A\rightarrow 1$ oscillations acquire specific triangular form.

The physical cause of the stability loss of $\pi$-mode, as well as other NNMs, is \emph{parametric resonance}. Let us consider this cause for the case of small vibrational amplitudes. For $A\rightarrow 0$ one can neglect all nonlinear terms in Eq.~\eqref{ved} and this equation then acquires the form
 \begin{equation}
 \ddot{V}+4V=0.
 \end{equation}
This is equation of the harmonic oscillator with eigenfrequency $\omega=2$ and, therefore, we can write its general solution in the form:
\begin{equation}\label{resh-pi-garm}
V(t)=A\sin(2t+\varphi_0),
\end{equation}
where $\varphi_0$ is an initial phase.

As we have already stated, the variational system for the Floquet analysis of $\pi$-mode can be decomposed into the set of independent equations~\eqref{pisimpl} with different values of the parameter $k$. As a consequence of the condition of small amplitudes, $|V(t)|\ll 1$ and one can decompose the function $c(t)=[1-V^2(t)]^{-1}$ into Taylor series. Taking into account only first nonlinear term, we obtain
\begin{equation}
\ddot{\delta}+4\gamma[1+V^2(t)]\delta=0,
\end{equation}
where $\gamma=\cos^2(\frac{k}{2})$. Introducing new time variable $\tau = 2t+\varphi_0$, allow us to transform this equation to the standard form of the Mathieu equation~\cite{abr}:
\begin{equation}\label{y_cos}
y''+\{a-2q\cos(2\tau)\}y=0.
\end{equation}
Here
\begin{equation}\label{aqrel}
y(\tau)=\delta(\frac{\tau-\varphi_0}{2}), ~a=(1+\frac{A^2}{2})\gamma, ~q=\frac{A^2}{4}\gamma,
\end{equation}
while differentiation with respect to the new time argument is denoted by prime.

It follows from Eq.~\eqref{aqrel} that there is a linear relation between parameters $a$ and $q$ of the Mathieu equation:
\begin{equation}\label{a-q-rel}
a=\gamma+2q.
\end{equation}
On the other hand, it is well known that there exists an infinite number of regions of unstable motion in the $(a-q)$ plane for the Mathieu equation~\eqref{y_cos} (see Ince-Strutt diagram in~\cite{abr}), and these regions appear because of parametric resonance.

\begin{figure}[htb]
\begin{center}
\includegraphics[scale=0.8]{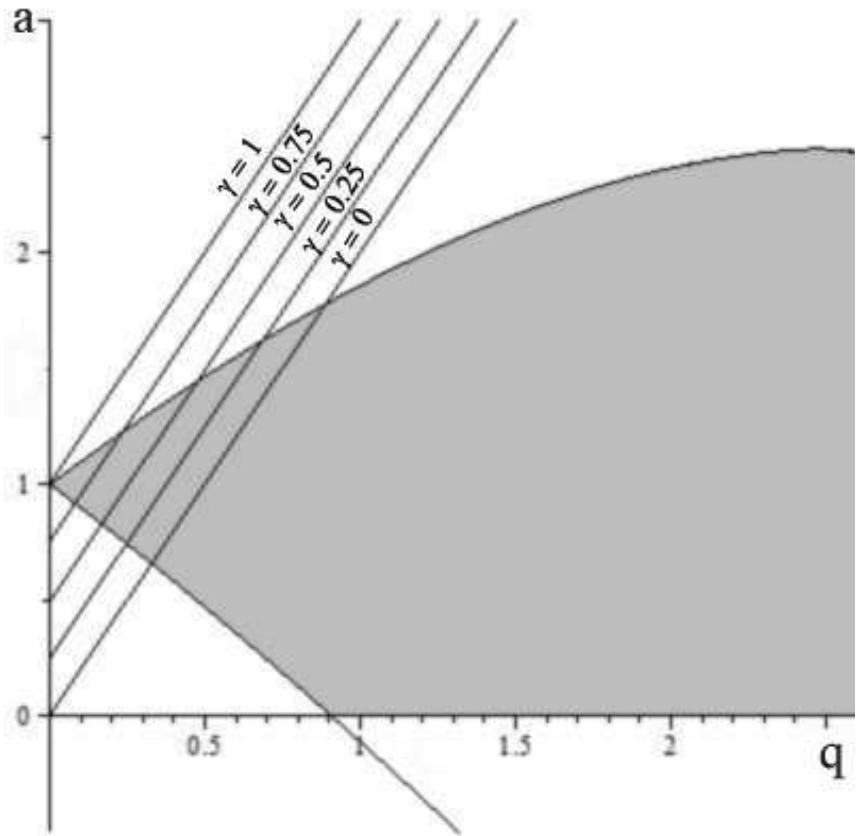}
\caption{First zone of instability for the Mathieu equation and a number of straight lines, determined by Eq.~\eqref{a-q-rel}}{\label{mat} }
\end{center}
\end{figure}

For studying stability loss of $\pi$-mode, it is sufficient to consider only the first region of instability from the Ince-Strutt diagram. We depict it by grey color in Fig.\ref{mat}. We also depict in this figure a number of straight lines $a=\gamma+2q$ [see Eq.~\eqref{a-q-rel}] for different values of the parameter $\gamma$, which is connected with the mode number $j$ by the relation
\begin{equation}\label{gamcos}
\gamma=\cos^2(\frac{\pi j}{N}).
\end{equation}

For $A=0$, the function $c(t)$ from Eq.~\eqref{capacity} turns out to be equal to unity and equation~\eqref{pisimpl} represents a set of harmonic oscillators with eigenfrequencies
\begin{equation}
\omega_j=\cos(\frac{k}{2})=\cos(\frac{\pi j}{N}),\; j=1..N.
\end{equation}
On the other hand, for $A\neq 0$, Eq.~\eqref{Pi_explicit_form} transforms to a set of the Mathieu equations with parameters $a$ and $q$ depending on the mode number $j$.

The function $c(t)$ is time-periodic and for certain mode numbers $j$ the coefficients $a(j)$ and $q(j)$ may hit the unstable region depicted in Fig.\ref{mat}. In such a case condition of the parametric resonance is satisfyed and, as a result, the corresponding $\delta_j(t)$ begin to grow exponentially. This means the loss of $\pi$-mode stability, because this mode can be stable only when all $\delta_j(t)$ continue to be small if their initial values $\delta_j(0) (j=1..N)$ are small.

Let us consider the interaction of $\pi$-mode with a "sleeping" linear normal mode $\delta_j(t)$, i.e. a mode equal to zero at the initial instant $t=0$. If we gradually increase the $\pi$-mode amplitude $A$ from the value $A=0$, the point $(a,q)$ moves upwards along one of the straight lines depicted in Fig.\ref{mat} from its initial position ($\gamma$, 0), because $a=\gamma+2q$ and $q=\gamma\frac{A^2}{4}$ [see Eqs.~\eqref{a-q-rel} and~\eqref{aqrel}]. The $\pi$-mode is stable up to intersection of the above straight line with \emph{lower boundary} of the grey region in Fig.\ref{mat}. This intersection corresponds to the maximal amplitude $A_c^j$ of $\pi$-mode for which excitation of the sleeping mode $\delta_j(t)$ still does not occur.

On the other hand, the value $A_c^j$ can be obtained as a solution of the equations
\begin{equation}\label{29}
   \begin{aligned}
   &a=1-q-\frac{q^2}{8}...,\\
   &a=\gamma+2q,
   \end{aligned}
\end{equation}
where the first equation defines the lower boundary of the gray region in Fig.\ref{mat} (see, for example,~\cite{abr}). As a result of solving Eqs.~\eqref{29}, we find\footnote{For $|q|\ll 1$ one can neglect the term $\frac{q^2}{8}$ in Eq.~\eqref{29}.}
\begin{equation}
q=\frac{1-\gamma}{3}
\end{equation}
Taking into account the relation $q=\gamma\frac{A^2}{4}$ and Eq.~\eqref{gamcos}, we obtain finally
\begin{equation}\label{Atg}
A^j_c=\frac{2}{\sqrt{3}}tg(\frac{\pi j}{N}).
\end{equation}

Obviously, the loss of $\pi$-mode stability is determined by that sleeping mode $\delta_j(t)$ for which minimal value of $A_c^j ~(j=1..N)$ corresponds. Then we conclude from Eq.~\eqref{Atg} that this minimal value $A_c$ corresponds to $j=1$, i.e. the mode $\delta_1(t)$ is excited firstly when $\pi$-mode amplitude $A$ begins to increases from zero. Hereafter, we call $A_c$ \emph{critical} $\pi$-mode amplitude because this mode loses its stability beyond $A_c$. It follows from Eq.~\eqref{Atg} that for $\pi$-mode $A_c \rightarrow 0$ when $N \rightarrow \infty$.

\begin{figure}[htb]
\begin{center}
\includegraphics[scale=0.5]{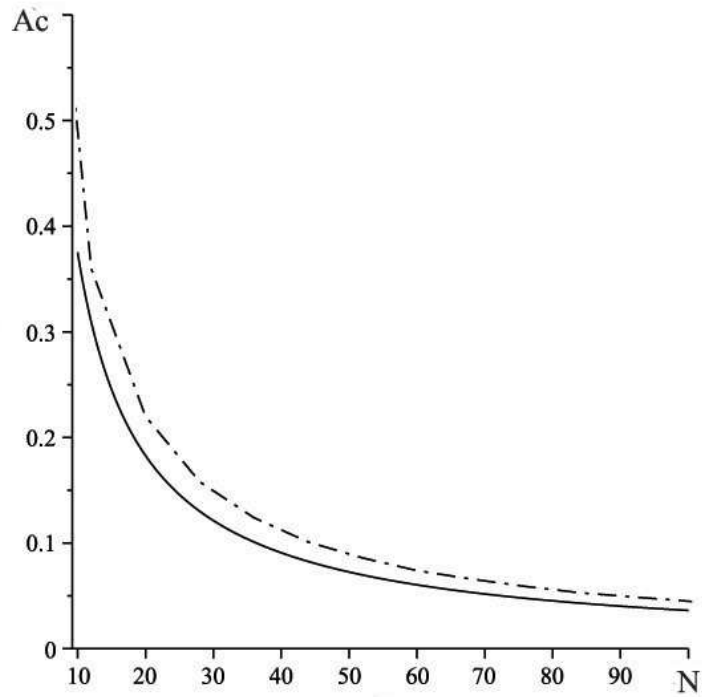}
\caption{Dependence of the critical amplitude $A_c$ on the number $N$ of the chain cells for $\pi$-mode. Solid line corresponds to the function $A_c(t)$ obtained from Eq.~\eqref{Atg} for $j=1$, while dot-dashed line was obtained as a result of numerical calculation}{\label{mat_ampl_pi} }
\end{center}
\end{figure}

In Fig.\ref{mat_ampl_pi}, we depict for $\pi$-mode the plot of the function $A_c(N)$ which shows dependence of the critical amplitude $A_c$ for different length ($N$) of the considered LC-chain. Solid line in Fig.\ref{mat_ampl_pi} corresponds to the function $A_c(N)$ obtained from Eq.~\eqref{Atg} for $j=1$, while dot-dashed line is depicted with the aid of the $\pi$-mode Floquet stability analysis. We calculate the Floquet exponents integrating Eq.~\eqref{Pi_explicit_form} for a given amplitude $V(0)=A$ and for all possible values of $\gamma$ detecting instability of $\pi$-mode if at least one of these exponents exceeds unity more than $\varepsilon=10^{-5}$.

Both plots in Fig.\ref{mat_ampl_pi} are in very good agreement for large values $N$. Some deviations taking place for small $N$ can be explained by approximation that was done for simplification of the governing equation~\eqref{ved}.

For $j=1$ and $N\gg1$, we find from Eq.~\eqref{Atg}
\begin{equation}\label{pi-as}
A_c(N)=\frac{2}{\sqrt{3}}tg\left(\frac{\pi}{N}\right)=\frac{2\pi}{\sqrt{3}}\frac{1}{N}+O\left(\frac{1}{N^3}\right).
\end{equation}
Therefore, decreasing of $A_c(N)$ for $N \rightarrow \infty$ is determined by the power law
\begin{equation}\label{Ac_st}
A_c(N)=CN^{-\beta},
\end{equation}
with $C\approx3.629, ~\beta=-1$.

Obviously, the above $\pi$-mode stability analysis based on the transition from Eq.~\eqref{Pi_explicit_form} to the Mathieu equation turns out to be \emph{exact} in the thermodynamic limit $N\rightarrow\infty$.
\section{Stability analysis of the nonlinear normal mode $\boldsymbol{\phi}_2$}\label{sec:6}

According to Eqs.~\eqref{SplitMatrix} and~\eqref{a2uM}, stability of NNM $\boldsymbol{\phi}_2$ is determined by stability of zero solution of the two-dimensional systems with different values $\gamma=e^{\frac{4\pi i}{N}j}$.

However, it is easy to note that matrices $\hat{A}, \hat{B}, \hat{D}$ from Eq. (\ref{a2uM}) \emph{commute} with each other and, therefore, they can be diagonalized \emph{simultaneously} with the aid of an appropriate unitary transformation. As a result of this transformation, we split each of these two-dimensional variational systems into two independent scalar differential equations.

Indeed, Eqs.~\eqref{SplitMatrix} and~\eqref{a2uM} can be written in the form:
\begin{equation}
\left\{
\begin{aligned}
&(1-V^2)\ddot{\delta}_1-4V\dot{V}\dot{\delta}_1-(2\dot{V}^2+V\ddot{V})\delta_1=-2\delta_1+(1+\gamma)\delta_2,\\
&(1-V^2)\ddot{\delta}_2-4V\dot{V}\dot{\delta}_2-(2\dot{V}^2+V\ddot{V})\delta_2=-2\delta_2+(1+\bar{\gamma})\delta_1.\\
\end{aligned}
\right.
\end{equation}
Introducing new infinitesimal variables $y_i=(1-V^2)\delta_i$ instead of the old variables $\delta_i ~(i=1,2)$ [see Eq.\eqref{y300}], we obtain
\begin{equation}\label{ymatr}
\begin{pmatrix} 1-V^2 & 0 \\ 0 & 1-V^2 \end{pmatrix}\ddot{\boldsymbol{y}}=\begin{pmatrix} -2 & 1+\gamma \\ 1+\bar{\gamma} & -2 \end{pmatrix}\boldsymbol{y},
\end{equation}
i.e.
\begin{equation}
\hat{M}\ddot{\boldsymbol{y}}=\hat{R}\boldsymbol{y}.
\end{equation}
With the aid of a certain  unitary transformation $\hat{T}$, one can diagonalize matrix $\hat{R}$ without changing matrix $\hat{M}$ which is proportional to the identical matrix. As a result of such transformation, we obtain:
\begin{equation}
\begin{pmatrix} 1-V^2 & 0 \\ 0 & 1-V^2 \end{pmatrix}\ddot{\boldsymbol{\epsilon}}=\begin{pmatrix} \lambda_1 & 0 \\ 0 & \lambda_2 \end{pmatrix}\boldsymbol{\varepsilon},
\end{equation}
where $\boldsymbol{\varepsilon}=\hat{T}\boldsymbol{y}$, while $\lambda_1=-2+2\sin(\frac{k}{2}), \lambda_2=-2-2\sin(\frac{k}{2})$ are eigenvalues of the matrix $\hat{R}$.

\begin{figure}[h]
\includegraphics[width=60mm]{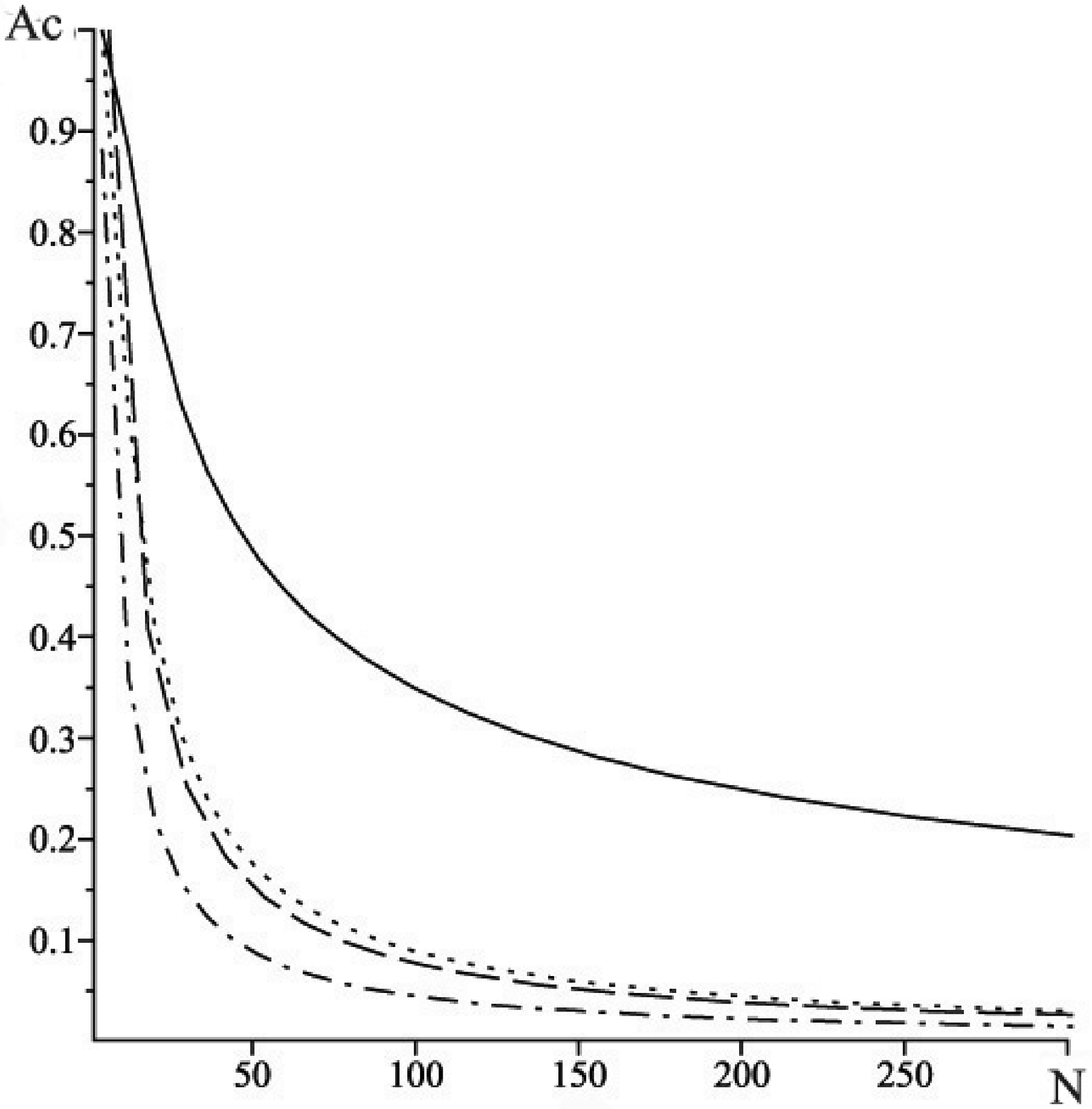}\hfill\includegraphics[width=60mm]{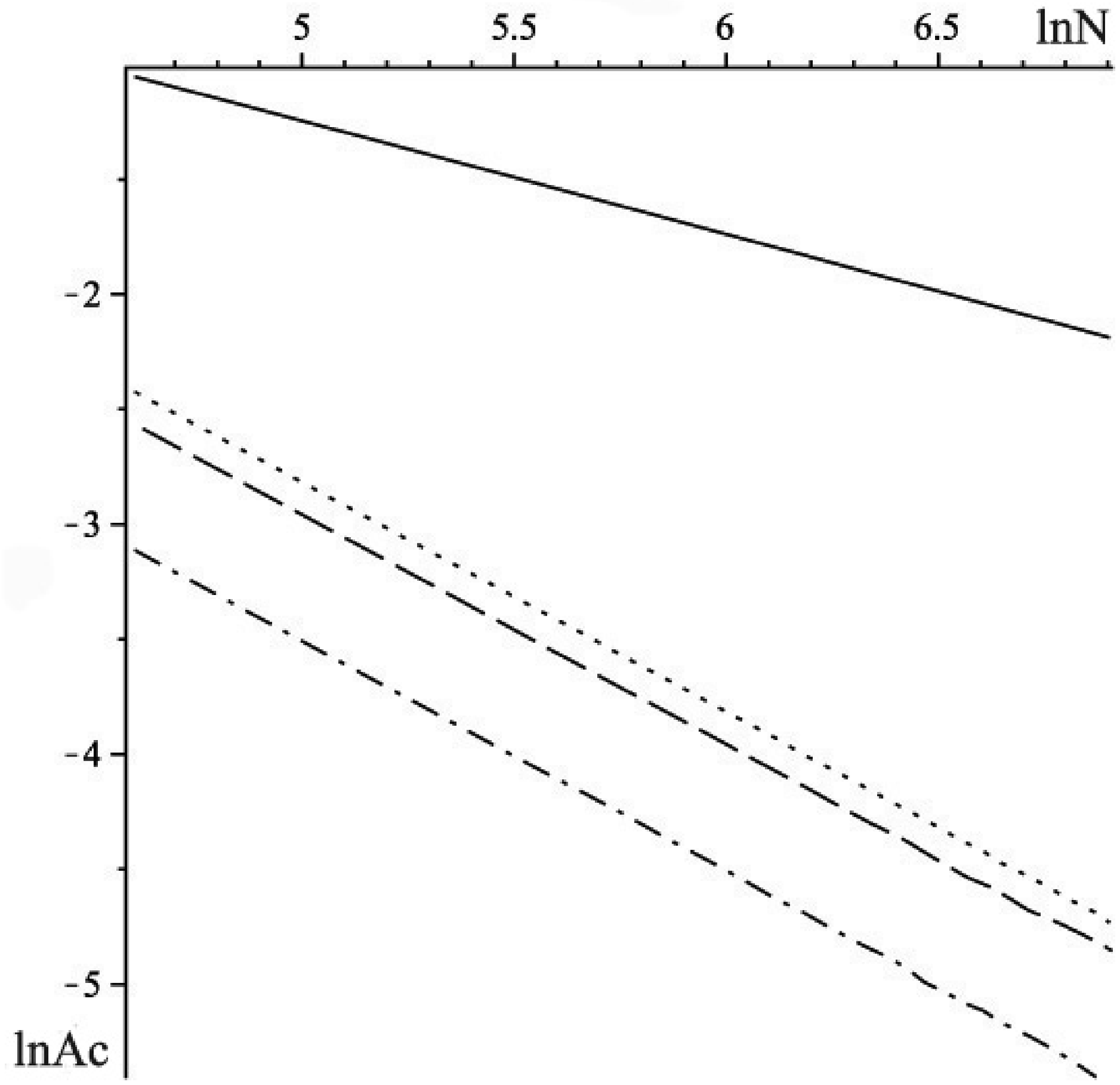}\\ \parbox{60mm}{\caption{Stability of nonlinear normal modes \eqref{pi_anzats}-\eqref{a6-anzats} for LC electrical chains with different $N$. Solid line corresponds to the mode $\boldsymbol{\phi}_2$ \eqref{a2u-anzats}, dashed line - to the mode $\boldsymbol{\phi}_3$ \eqref{a4-anzats}, dot-dashed line - to the $\pi$-mode  $\boldsymbol{\phi}_1$ \eqref{pi_anzats}, dotted line - to the modes  $\boldsymbol{\phi}_4$ \eqref{a3-anzats} and $\boldsymbol{\phi}_5$ \eqref{a6-anzats}}\label{fig3}}
\hfill\parbox{60mm}{\caption{Plots of $A_c(N)$ for nonlinear normal modes \eqref{pi_anzats}-\eqref{a6-anzats} in logarithmic scale. Solid line corresponds to the mode $\boldsymbol{\phi}_2$ \eqref{a2u-anzats}, dashed line - to the mode $\boldsymbol{\phi}_3$ \eqref{a4-anzats}, dot-dashed line - to the $\pi$-mode  $\boldsymbol{\phi}_1$ \eqref{pi_anzats}, dotted line - to the modes  $\boldsymbol{\phi}_4$ \eqref{a3-anzats} and $\boldsymbol{\phi}_5$ \eqref{a6-anzats}}\label{ln}}
\end{figure}

Thus, stability analysis of the mode $\boldsymbol{\phi}_2$ is reduced to studying stability of zero solutions of the independent equations:
\begin{equation}\label{ysin}
\left\{
\begin{aligned}
&\ddot{\varepsilon}_1+2c(t)[1-\sin(\frac{k}{2})]\varepsilon_1=0,\\
&\ddot{\varepsilon}_2+2c(t)[1+\sin(\frac{k}{2})]\varepsilon_2=0,\\
\end{aligned}
\right.
\end{equation}
where $c(t)=(1-V^2)^{-1}$. It is obvious, that these equations can be investigated in the thermodynamic limit $N\rightarrow\infty$ in the same way as the variational equation~\eqref{pisimpl} for $\pi$-mode (note, that $k=\frac{2\pi}{N}j$ for $\pi$-mode, while in the case of $\boldsymbol{\phi}_2$-mode, $k=\frac{4\pi}{N}j$). The corresponding result read:
\begin{equation}
A_c(N)=\frac{2}{\sqrt{3}}\sqrt{\frac{\sin(\frac{2\pi}{N}j)}{1-\sin(\frac{2\pi}{N}j)}}.
\end{equation}
The scaling of $A_c(N)$ in the limit $N\rightarrow\infty$ for the mode $\boldsymbol{\phi}_2$, as well as that for the mode $\boldsymbol{\phi}_1$, is determined by Eq.~\eqref{Ac_st}. However, in this case, the exponent $\beta$ is equal $\frac{1}{2}$. The final stability result for NNM $\boldsymbol{\phi}_2$ is depicted in Fig.\ref{fig3}.

We have reached the complete splitting of the variational system for NNM $\boldsymbol{\phi}_2$ in two steps. Firstly, we have used the general group-theoretical approach based on the symmetry group $G=[\hat{a}^2\hat{u}]$ and obtained two-dimensional system~\eqref{ymatr}. Secondly, we have split these system into individual scalar equations~\eqref{ysin}. Let us note that one can obtain this final result in one step only, if splitting is produced on the base of the full symmetry group $G_2=[\hat{a}^4,\hat{\imath}\hat{u}]$ of the mode $\boldsymbol{\phi}_2$.

\section{Stability analysis of NNM $\boldsymbol{\phi}_3$}\label{Sec:7}

Stability analysis of nonlinear normal modes $\boldsymbol{\phi}_3$ is reduced to studying two-dimensional system, determined by Eqs.\eqref{SplitMatrix}, \eqref{a4M}, which \emph{cannot} be split into scalar equations. Indeed, matrices $\hat{A}$ and $\hat{B}$ don't commute with the matrix $\hat{D}: \hat{A}\hat{D}\neq\hat{D}\hat{A}, \hat{B}\hat{D}\neq\hat{D}\hat{B}$. As a consequence, it is impossible to diagonalize these three matrices, $\hat{A}, \hat{B}, \hat{D}$, simultaneously, i.e. with the aid of one and the same unitary transformation, as it was possible in the case of the mode $\boldsymbol{\phi}_2$. In turn, this means that two-dimensional variational subsystem for NNM $\boldsymbol{\phi}_3$ cannot be decomposed into independent scalar equations and we are forced to study zero solution stability for the following 2D system:
\begin{equation}
\left\{
\begin{aligned}
&c(t)\ddot{\delta}_1+g(t)\dot{\delta}_1-f(t)\delta_1-(1+\gamma)\delta_2=0,\\
&\ddot{\delta}_2-2\delta_2-(1+\bar{\gamma})\delta_1=0,\\
\end{aligned}
\right.
\end{equation}
where $\gamma=e^{ik} (k=\frac{4\pi j}{N})$. The time-dependent functions $c(t), g(t)$ and $f(t)$ are defined in Eq.~\eqref{cfg_def}.

We can consider infinitesimal variables $\delta_1$ and $\delta_2$ as linear normal modes (phonon modes) corresponding to certain wave numbers. Let us remind, that, rigorously speaking, every phonon mode represents time-dependent variable $\delta_k(t)$ multiplied by a certain $N$-dimensional vector determining a distribution of voltages on all capacitors of the electrical chain. We call $\delta_k(t)$ "phonon mode" only for brevity. It follows from the matrix $S$, which leads to splitting of the original variational systems, that the above normal modes, $\delta_1$ and $\delta_2$, are associated with different wave numbers, $k$ and $\pi-k$:
\begin{equation*}
\delta_1=\delta_k(t), \delta_2=\delta_{\pi-k}(t).
\end{equation*}

It is interesting to note that the wave numbers $k$ and $\pi-k$ situated symmetrically with respect  to the center of primitive cell $[0;\pi]$ of the reciprocal lattice which corresponds to the original lattice of our electrical chain in vibrational state describing by the mode $\boldsymbol{\phi}_3$\footnote{The primitive cell $[0;\frac{2\pi}{a}]$ in the reciprocal lattice corresponds to the one-dimensional lattice with period $a$.}.

As in the case of $\boldsymbol{\phi}_1$- and $\boldsymbol{\phi}_2$-modes, NNM $\boldsymbol{\phi}_3$ turns out to be stable for sufficiently small amplitudes $A=V(0)$. Increasing of the amplitude leads to parametric resonance appearance which results in exciting two sleeping phonon modes $\delta_1=\delta_k(t)$ and $\delta_2=\delta_{\pi-k}(t)$ \emph{simultaneously}.

The stability loss of NNM $\boldsymbol{\phi}_3$ is determined by those sleeping phonon modes (the modes with certain numbers $j$) which are firstly excited as a result of increasing  the amplitude of this nonlinear normal mode. Calculating the Floquet exponents, for electrical chains for different $N$, we can detect the corresponding phonon modes. In turn, this give us a possibility to obtain the function $A_c(N)$ similar to that depicted in Fig.\ref{mat_ampl_pi} for the mode $\boldsymbol{\phi}_1$.

\section{Stability analysis of modes $\boldsymbol{\phi}_4$ and $\boldsymbol{\phi}_5$}\label{sec:8}
The stability analysis of nonlinear normal modes $\boldsymbol{\phi}_4$ and $\boldsymbol{\phi}_5$ is reduced to studying zero solution stability of the one and the same three-dimensional system determined by Eqs.~\eqref{SplitMatrix} and~\eqref{a3M}:
\begin{equation}
\left\{
\begin{aligned}
&c(t)\ddot{\delta}_1+g(t)\dot{\delta}_1-f(t)\delta_1-\delta_2-\gamma\delta_3,\\
&\ddot{\delta}_2-\delta_1-2\delta_2-\delta_3=0,\\
&c(t)\ddot{\delta}_3+g(t)\dot{\delta}_3-\bar{\gamma}\delta_1 -\delta_2-\delta_3=0.\\
\end{aligned}
\right.
\end{equation}
Here the functions $c(t), g(t), f(t)$ are given in Eq.~\eqref{cfg_def}, while $\gamma=e^{ik}$ with $k=\frac{6\pi j}{N} (j=1,2,...,\frac{N}{6})$.

Results on stability of NNMs $\boldsymbol{\phi}_4$, $\boldsymbol{\phi}_5$ are presented in Fig.\ref{fig3}. It is interesting to note that the stability plots for the modes $\boldsymbol{\phi}_4$ and $\boldsymbol{\phi}_5$ turn out to be \emph{identical}, despite the different functions $V(t)$, determining the coefficients $c(t), g(t), f(t)$, correspond to these modes. Indeed, the functions $V(t)$ for the modes $\boldsymbol{\phi}_4$ and $\boldsymbol{\phi}_5$ are obtained from the governing equation (\ref{ved})  with $\mu=3$ and $\mu=1$, respectively.

\section{Asymptotic behavior of $A_c(N)$ for $N\rightarrow\infty$}\label{sec:9}

The critical amplitudes $A_c(N)$ for all symmetry-determined nonlinear normal modes~\eqref{pi_anzats}-\eqref{a6-anzats} \emph{tend to zero} when $N \rightarrow \infty$. For the mode $\boldsymbol{\phi}_1$ and $\boldsymbol{\phi}_2$ the corresponding scaling law of decreasing $A_c(N)$ was presented in Eq.~\eqref{Ac_st} with $\beta$ equal to $1$ and $\frac{1}{2}$, respectively. Naturally, it is interesting to find similar asymptotic formulas for other NNMs. Numerical experiments allow us to obtain Fig.\ref{ln} where we present functions $A_c(N)$ in logarithmic scale for large $N ~(N>100)$. It is obvious from the plots depicted in this figure that $A_c(N)$ for all NNMs are \emph{power functions} since dependence of $\ln A_c(N)$ on $\ln N$ represents straight lines. The coefficient $\beta$ in the formula $A_c(N)=CN^{-\beta}$ is determined by inclination of the corresponding straight line to the horizontal axis.

In this way, we have obtained the following values of the constants $\beta$ and $C$ entering the law $A_c(N)=CN^{-\beta}$:
\begin{equation}\label{scal_law}
\begin{aligned}
&\boldsymbol{\phi}_1 [\hat{a}^2, \hat{\imath}\hat{u}] ~\text{--- $\pi$-mode [Eq.~\eqref{pi_anzats}]:}~\beta_1=0.99, C_1=3.62;\\
&\boldsymbol{\phi}_2[\hat{a}^4, \hat{\imath}\hat{u}] ~\text{[Eq.~\eqref{a2u-anzats}]:}
~\beta_2=0.49, C_2=2.89;\\
&\boldsymbol{\phi}_3 [\hat{a}^4, \hat{a}\hat{\imath}] ~\text{[Eq.~\eqref{a4-anzats}]:}~\beta_3=0.99, C_3=8.57;\\
&\boldsymbol{\phi}_4 [\hat{a}^3, \hat{\imath}\hat{u}], \boldsymbol{\phi}_5 [\hat{a}^3\hat{u}, \hat{a}\hat{\imath}\hat{u}] ~\text{[Eqs.~\eqref{a3-anzats} and~\eqref{a6-anzats}]}:~\beta_4=0.99, C_4=7.54.
\end{aligned}
\end{equation}

Relying on the above data, we can assert that for all NNMs, except for the mode $\boldsymbol{\phi}_2$, the constant $\beta$, determining the rate of decreasing of the critical amplitude $A_c$ with increasing $N$, is equal to unity ($\beta=1$), while for $\boldsymbol{\phi}_2$ it is equal to $\frac{1}{2}$. Note that we have obtained the values of the parameter $\beta$ for the modes $\boldsymbol{\phi}_1$ and $\boldsymbol{\phi}_2$ analytically, as well as numerically, while for NNMs $\boldsymbol{\phi}_3, \boldsymbol{\phi}_4, \boldsymbol{\phi}_5$ this parameter was found only numerically. We conjecture that it can be obtained also analytically by the specific asymptotic method developed in~\cite{ryabov-2011}. However, this was not done in the present paper, because numerical results~\eqref{scal_law} seem to be sufficient for hypothesis that the parameter $\beta$ for NNMs $\boldsymbol{\phi}_3, \boldsymbol{\phi}_4$ and $\boldsymbol{\phi}_5$ is equal to unity. Indeed, the numerical data~\eqref{scal_law} show that the deviation of the parameter $\beta$ from $1$ for modes $\boldsymbol{\phi}_1,\boldsymbol{\phi}_3, \boldsymbol{\phi}_4, \boldsymbol{\phi}_5$ and that from $\frac{1}{2}$ for the mode $\boldsymbol{\phi}_2$ are equal to $0.01$. We associate these deviations with some computational errors, because numerical data demonstrate the \emph{same} deviation from the exact values $\beta_1=1, \beta_2=\frac{1}{2}$ which were found in Sects.~\ref{sec:5},\ref{sec:6}. Naturally, we can assume that exact values of $\beta_3,\beta_4$ and $\beta_5$ are equal to unity.

\section{Studying of nonlinear normal modes stability in electrical chains with resistance and harmonic external voltage sources}\label{sec:10}

In the previous sections of this paper, we have studied existence and stability of nonlinear normal modes in the LC-chain considered in Ref.~\cite{Osting}. However, this model is rather idealized because it does not take into account the resistance of the electrical circuit. This resistance leads to dissipation of the initial excitation energy, and as a consequence, to impossibility of existence of undamped oscillations. Below, we introduced a generalized model, which we call LCR-chain. In each cell of this model there is an external source of time-periodic voltage. To exist the NNMs, described by Eqs.~\eqref{pi_anzats}-\eqref{a6-anzats}, the certain distribution of the above voltage sources along the chain is necessary.

\begin{figure}[htb]
\begin{center}
\includegraphics[scale=0.75]{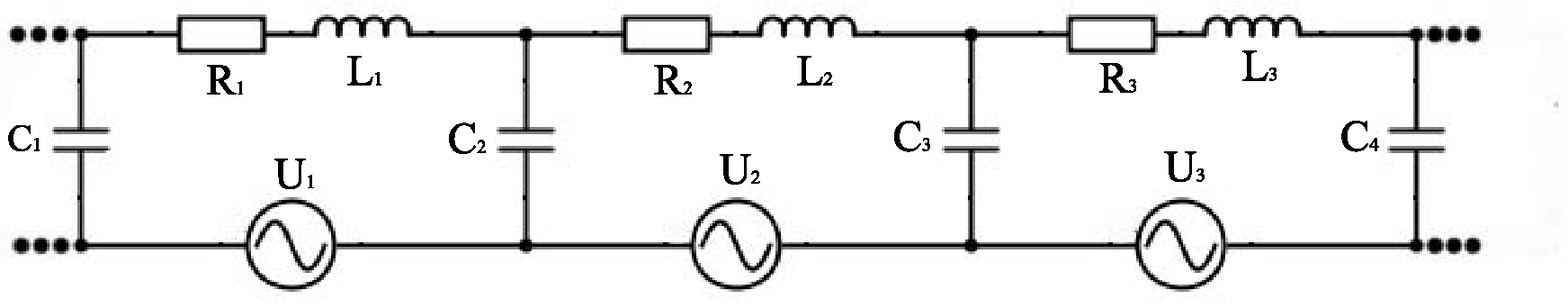}
\caption{LCR-chain}\label{scheme2}
\end{center}
\end{figure}

The goal of our further discussion is to demonstrate that the group-theoretical methods, developed in Sects.~\ref{sec:2}-\ref{sec:9} for studying stability of NNMs in LC-chains, are suitable for the similar analysis of these dynamical objects in the LCR-chains. The chain of the latter type (see Fig.\ref{scheme2}) represents a set of cells, consisting of capacitors ($C_j$), inductors ($L_j$), resistors ($R_j$) and the harmonic external sources of voltage [$U(t)=U_j\sin(\Omega t)$].

Periodic boundary conditions are assumed, as was done in our studying of LC-chains. All inductors, capacitors and resistors are considered to be equal ($L_j=L_0,C_j=C_0, j=1..N$) and, therefore, electrical cells differ from each other only by amplitudes $U_j$ of the external voltages.

Applying Kirchhoff's laws, we obtain the following equations for dynamics of the LCR-chain:
\begin{subequations}\label{kir'}
   \begin{eqnarray}
   &L_0\frac{dI_j}{dt}=V_j-V_{j+1}-I_j R+U_j\cos(\Omega t),\\
   &C_0(1-V_j^2)\frac{d V_j}{dt}=I_{j-1} - I_j.
   \end{eqnarray}
\end{subequations}
Here $V_j=V_j(t)$ is the voltage on the $j$-th capacitor, while $I_j=I_j(t)$ represents the current through the $j$-th resistor and the corresponding inductor. The appropriate scaling of variables in Eq.~\eqref{kir1} allows one to suppose $L_0 = 1$ and $C_0 = 1$:
\begin{subequations}\label{kir1}
   \begin{eqnarray}
   &\frac{dI_j}{dt}=V_j-V_{j+1}-I_j r+U_j\cos(\Omega t),\\
   &(1-V_j^2)\frac{d V_j}{dt}=I_{j-1} - I_j,
   \end{eqnarray}
\end{subequations}
where $r=R\sqrt{C_0/L_0}$.

In Sec.2, we have found that for LC-chains with the appropriate number of cells ($N$) only five symmetry-determined nonlinear normal modes can exist. Each of these modes is invariant with respect to a subgroup of the symmetry group (group of invariance) of the dynamical equations ~\eqref{model_with_bound} of the considered model. It is easy to understand that a given nonlinear normal mode can exist in the LCR-chains only if the distribution of the external voltages $U_j$ possesses the symmetry group equal to that of the chosen NNM.

Below, we consider the stability of $\pi$-mode to exemplify the application of our group-theoretical methods for the case of electrical chains with dissipation and external voltages.

For existence of $\pi$-mode, i.e. the dynamical regime of the form
\begin{equation}\label{pi_dist}
\{V(t),-V(t)|V(t),-V(t)|V(t),-V(t)|\ldots\},
\end{equation}
it requires that the external voltage distribution has the same space symmetry as this mode:
\begin{equation}\label{U_pi_dist}
\{U(t),-U(t)|U(t),-U(t)|U(t),-U(t)|\ldots\}.
\end{equation}
The substitution Eqs.~\eqref{pi_dist},\eqref{U_pi_dist} into Eqs.~\eqref{kir1}, reduces these $2N$ first-order equations to a system of two equations of the form:
\begin{subequations}\label{pi_syst}
   \begin{eqnarray}
    &\frac{dI}{dt}=2V-Ir+U_0\cos(\Omega t),\\
   &(1-V^2)\frac{dV}{dt}=-2I.
   \end{eqnarray}
\end{subequations}
All other equations of our dynamical model ~\eqref{kir1} are equivalent to Eqs. ~\eqref{pi_syst}.

It is important to note that the $\pi$-mode ~\eqref{pi_dist} for LCR-chain, in contrast to the similar dynamical object for LC-chain, is no longer periodic. Indeed, the considered dynamical regime in the LCR-chain represents a superposition of natural oscillations with frequency $\omega = 2$ (which occur for the case $R=0, V_0=0$) and the forced vibrations with frequency $\Omega$ of external sources. After a transient, the natural oscillations are damped because of energy dissipation on resistors, while vibrations with frequency $\Omega$ are survived.

Now we show that our group-theoretical method for studying stability can be used for splitting high-degrees variational system into independent subsystems not only for periodic dynamical regime, but also for non-periodic transient. Without repeating the lengthy calculations given for the LC-chain in Sects.~\ref{sec:4}-\ref{sec:5}, we present, for the LCR-chain, only the final results.

$2N$-dimensional variational system obtained by linearization of Eqs.~\eqref{kir1} in the vicinity of the $\pi$-mode ~\eqref{pi_dist} is split into $N$ independent second-order equations as follows:
\begin{equation}\label{ras-z}
\ddot{z}+r\dot{z}+\frac{4\cos^2{q}}{1-V^2(t)}z=0
\end{equation}
Here $z(t)$ represents voltage infinitesimal perturbation, $q = \frac{\pi j}{N} (j = 1 .. N)$ is the wave number corresponding to a plane wave, while $V(t)$ is determined by Eqs. ~\eqref{pi_syst}. Thus, we actually consider the stability of $\pi$-mode ~\eqref{pi_dist} with respect to infinitesimal perturbations in the form of individual plane waves.

Let us emphasize that equation ~\eqref{ras-z} are valid not only for the periodic oscillations of the voltage $V(t)$, but also for arbitrary dynamical regimes with the spatial structure ~\eqref{pi_dist}. This means, that our group-theoretic method for splitting variational systems allows one to analyze the stability of dynamical regimes even in the cases where the standard Floquet method cannot be applied.

The complete stability analysis of the $\pi$-mode in the LCR-chain is extremely cumbersome, because its stability properties are different in the different domains of parameters $U_0$, $R$ and $\Omega$. For this reason, we analyze here the stability of $\pi$-mode for only one set of these parameters to illustrate the efficiency of the group-theoretic method developed in the present paper to simplify the stability studying of the dynamic regimes in systems with discrete symmetry.
We consider the following values of the LCR-chain parameters:
\begin{equation}\label{set-par}
R = 0.01, \Omega = 2.2, U_0 = 0.084, N = 20.
\end{equation}
Note, that the frequency $\Omega=2.2$ is close enough to the natural frequency $\omega=2$ of the LC-chain, i.e. for the case $R=0, U_0=0$.

\begin{figure*}[htb]
\includegraphics[scale=0.25]{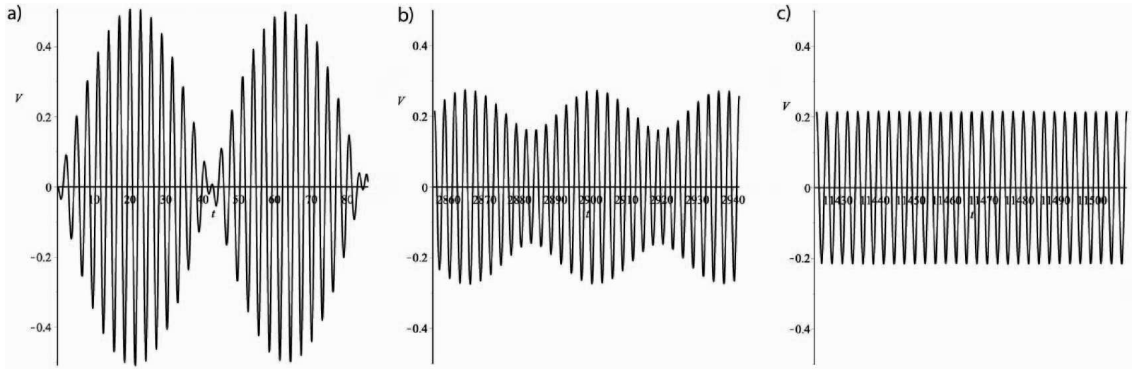}
\caption{Time evolution of the voltage $V(t)$ for the set of the LCR-chain parameters~\eqref{set-par}. \newline
a) $t=0..30T$, b) $t=1000T..1030T$, c) $t=4000T..4030T$, where $T=\frac{2\pi}{\Omega}$}
\label{V_evol}
\end{figure*}

In Fig.\ref{V_evol}, we show the time evolution of $V(t)$ for the above set of the LCR-chain parameters. From this figure, one can see a transient followed by the stationary regime with constant frequency and amplitude. The parameters of the stationary oscillations can be found by the RWA-method applied to the $2\times2$ system ~\eqref{pi_syst}.

In accordance with the RWA-approximation, we look for the voltage $V(t)$ and the current $I(t)$ in the form
\begin{equation*}
V(t)=a\cos{\Omega t}+b\sin{\Omega t}, I(t)=A\cos{\Omega t}+B\sin{\Omega t}
\end{equation*}
with arbitrary constants $a, b, A$ and $B$. Substituting these expressions into Eqs.~\eqref{pi_syst} and omitting the terms with tripled frequency ($3\Omega$), we obtain
\begin{equation*}
a=0.21, b=0, A=0 , B=0.23.
\end{equation*}
These values are in good agreement with the results of numerical experiments presented in Fig.\ref{V_evol}c. It is essential that the oscillation amplitude in some time intervals of the transient exceed significantly the amplitude of stationary regime. As a result, the dynamical regime ~\eqref{pi_dist} losses stability \emph{already during transient}.

Analysis of the equation ~\eqref{ras-z} for different $q$ gives stability thresholds for the infinitesimal perturbations in the form of plane waves with different wave numbers. The stability loss threshold is determined by the modes with the longest wavelengths, i.e. those with the smallest wavenumbers $k=\frac{2\pi q}{N} (q \sim 1)$.

Let us note that at the transient stage the electrical vibrations in the LCR-chain are not periodic, and therefore the standard Floquet method can not be applied. On the other hand, our group-theoretical method allows one to split $2N$-dimensional variational system into $N$ independent two-dumensional subsystems. This makes it possible to greatly simplify the stability analysis of $\pi$-mode in chains of arbitrary length, in particular, for the case $N\rightarrow\infty$.

\section{Conclusion}
In the present paper, we study periodic vibrations in the chain of nonlinear capacitors coupled by linear inductors (LC-chain), assuming that voltage-dependence of capacitors is described by the function $C(V)=C_0(1-bV^2)$ as was done in Ref.~\cite{Osting}. These dynamical regimes represent symmetry-determined nonlinear normal modes and we prove that only five types of such modes can exist in the considered electrical chains. The stability analysis of these nonlinear modes with the aid of the straightforward Floquet method turns out to be very difficult in the case of large number ($N$) of the electrical chain cells.

We apply the general group-theoretical method, developed in~\cite{zh-PhRevE}, for a radical simplification of the stability studying. This method allows us to reduce large-dimensional variational systems, appearing in the stability analysis of nonlinear normal modes, to independent subsystems of small dimensions. Even in the case $N\rightarrow \infty$ for all such modes it is sufficient to study zero solution stability of one-, two-, and three-dimensional variational systems. Using this approach, we were able to find the voltage amplitude $A_c(N)$ of a given NNM which corresponds to onset of unstable vibrations in the electrical chain for any fixed $N$. Moreover, we justify the scaling law $A_c(N)=CN^{-\beta}$ with parameter $\beta$ equal to $\frac{1}{2}$ for the mode $\boldsymbol{\phi}_2$ [see Eq.~\eqref{a2u-anzats}] and equal to unity for all other NNMs.

It is essential that our group-theoretical method can be applied not only to stability studying of periodic dynamical regimes, but also to other regimes. In particular, this is true for \emph{bushes} of nonlinear normal modes which represent quasiperiodic vibrations~\cite{DAN-1,DAN-2,PHD-1998}. In such a case, one cannot apply the standard Floquet method. However the variational systems corresponding to these dynamical objects can be also decomposed into subsystems of sufficiently small dimensions and this simplifies the stability analysis. For the case of electrical chain~\eqref{model_with_bound}, this problem will be considered in a future paper.

In the last section, we study the nonlinear normal modes and their stability in the LCR-chains, containing not only inductors and capacitors, but also resistors and external sources of time-periodic voltage. We show how our group-theoretical method can be used in this more general case.

\end{document}